\documentclass[jcp,aip,amsmath,amssymb,amsfonts,groupedaddress]{revtex4-1}
\usepackage{graphicx}
\usepackage{color}

\newcommand{\be}{\begin{equation}}
\newcommand{\ee}{\end{equation}}
\newcommand{\bea}{\begin{eqnarray}}
\newcommand{\eea}{\end{eqnarray}}

\begin{document}

\title
{A simple closure procedure for the study of velocity
autocorrelation functions in fluids as a ``bridge'' between
different theoretical approaches}
\author{V.V.~Ignatyuk}
\email[Corresponding author:~]{ignat@icmp.lviv.ua}
\author{I.M.~Mryglod}
\author{T.~Bryk}
\affiliation{Institute for Condensed Matter Physics, 1
Svientsitskii Street, 79011, Lviv, Ukraine}

\begin{abstract}
Velocity autocorrelation functions (VAF) of the fluids are studied
on short- and long-time scales within a unified approach. This
approach is based on an effective summation of the infinite
continued fraction at a reasonable assumption about convergence of
relaxation times of the high order memory functions, which have
purely kinetic origin. The VAFs obtained within our method are
compared with computer simulation data for the liquid Ne at
different densities and the results, which follow from the
Markovian approximation for the highest order kinetic kernels. It
is shown that in all the thermodynamic points and at the chosen
level of the hierarchy, our results agree much better with the MD
data than those of the Markovian approximation. The density
dependence of the transition time, needed for the fluid to attain
the hydrodynamic stage of evolution, is evaluated. The common and
distinctive features of our method are discussed in their
relations to the generalized collective mode (GCM) theory, the
mode coupling theory (MCT), and some other theoretical approaches.

\end{abstract}

\maketitle

\section{Introduction\label{Introduct}}

Dynamics of many-body systems at various time scales remains as
one of the main problems of the non-equilibrium statistical
mechanics despite its long history and considerable achievements
obtained so far. In the historical retrospective, one should first
mention the method of kinetic equations. Though initially this
method was designated for the study of the system dynamics,
arising from the uncorrelated collisions between two particles in
diluted gases \cite{kin-eq-gas}, its subsequent generalizations
allow one to describe more realistic systems like dense gases and
plasmas \cite{BGK}.

Nevertheless, it has to be stated that systematic studies of the
time behavior of many-particle systems started with the pioneering
work by N.N.~Bogoluyubov \cite{Bogolyubov1}, where the concept of
weakening correlations allowed one to reformulate a completely
unsolvable problem of the $N$-body dynamics into a much more
manageable task in terms of the correlation functions of lower
orders $s\ll N$.

Within such an approach, a time hierarchy of the phy\-si\-cal
stages, through which the system passed during its evolution
toward equilibrium (starting from the kinetic stage via
intermediate steps till the hydrodynamic one and beyond
\cite{MorozovBook}), appears quite naturally. The corresponding
time scales within that hierarchy can serve as a basis for
generalized hydrodynamics, and in particular in the generalized
collective mode theory (GCM) \cite{GCM1,GCM2}, where the set of
dynamical variables for description of long- and short-time
correlations usually consists of the densities of conserved
quantities and their time derivatives up to a certain order. In
the framework of the GCM, one actually comes to a simple
mathematical problem, expressing collective processes in the
system in terms of the dynamic eigenmodes: complex (real)
eigenvalues for propagating (relaxing) processes and corresponding
eigenvectors which characterize contribution of a particular
eigenmode to relevant excitations. The definite
advantages of the GCM consist in the facts that: i) the theory
describes correctly all the spectrum of collective excitations for
different kinds of fluids \cite{GCM3,GCM4,MryglodFolk,PRE_polar};
ii) the theory is a computer adapted one: all the elements of the
generalized dynamic matrix could be expressed via the static
correlation functions (SCF) and corresponding relaxation times,
which can be obtained by molecular dynamic simulations (MD)
\cite{GCM2,GCM3,GCM4}; iii) the time correlation functions (TCF)
calculated in the framework of the GCM obey the corresponding sum
rules (i.e. time expansion series reproduce the results of the
exact dynamics up to a ceratin order).

There is also another method, bearing a close resemblance with the
GCM. Using the continued fraction representation
\cite{Bafile1,Bafile2,Bafile3} for the Laplace transforms of the
VAFs, one can present the corresponding autocorrelation functions
(as well as the kinetic kernels of higher orders) as a weighted
sum of the exponents. This method operates with an
orthogonalized set of the dynamical variables, and the input
parameters in such an approach are the SCFs up to the $s$-th
order, while hydrodynamic correlation times (the zeroth time
moments of the corresponding TCFs), which are inherent to the GCM,
are not considered within such a scheme. Though, formally, one can
construct the infinite continued fraction, for practical purposes
it is necessary to truncate the fraction at the certain level
(say, the $s$-th) of the hierarchy. Such an ansatz is related to
the Markovian approximation \cite{GCM3} for the highest order
kinetic kernels (MA), and will be also used in our present paper.
Since in both cases the generalized dynamic matrixes are
constructed starting from the same conceptual points, there are
strong reasons to believe that the results for the TCFs, obtained
in the framework of the GCM and within the MA, at the chosen
hierarchy level, are close to each other to a high degree of
accuracy. This allows one to use the term ``mode approximation''
for the results obtained within the MAs.

Although the GCM can be extended by taking into account the
``ultraslow'' processes (defined by the time integrals of
corresponding densities \cite{Ome00,GCM_slow,Bry10}), which couple
in local approximation with hydrodynamic and extended dynamic
variables, usually the problems of account for slow structural
relaxation \cite{stuctur_relax} are approached by theories with
non-local coupling of dynamic variables. In the framework of the
mode coupling theory (MCT) \cite{MCT1,MCT2}, a basic set of the
dynamic variables consists of higher products \cite{MCT3} rather
than higher derivatives of the densities of conserved quantities.
Like in the GCM, a time-spatial dispersion of the kinetic kernels
gives rise to some peculiarities of TCF. However, in contrast to
the GCM, where the system description at long times is a
``bottleneck'' of the theory, the MCT approach mani\-fests its
efficiency upon the hydrodynamic stage of the system evolution
yielding $t^{-3/2}$ power law for the TCFs \cite{long_tail1}.
In its turn, the MCT (at least, in its standard
formulation \cite{MCT1}) cannot reproduce the sum rules, being
unreliable at the description of the fluids dynamics at the short
and intermediate times.

Afterwards, a lot of attempts have been made to
generalized the MCT, starting from the fundamental works by
G\"otze and co-workers \cite{Goetze1,Goetze2,Goetze3} up to more
recent papers \cite{Szamel,Reichman1}. In particular, the main
idea of the paper by Szamel \cite{Szamel} was not to factorize
immediately the memory function expressions leading to the
ordinary MCT, but to write the equation of the next hierarchy
level, and then factorize the memory function for the new
equation. This approach allowed to extend the predictions of the
ergodicity breaking transition to higher density cases. In
Ref.~\cite{Reichman1} the approach based on the truncation of the
coupled set of integro-differential equations for the 2n-point
density correlators at high enough level $s=1000$ gave the authors
possibility to describe quite perfectly a lot of peculiarities of
the ``density--density'' TCFs depending on the generalized bar
frequencies and coupling constants, and to put ahead new insights
for kinetic theories of the glass transitions.

Very recently, a microscopic interpretation of the long-time tails
of VAF was suggested from analysis of its memory kernels
\cite{MHforMemKer}. It was shown that the hydrodynamically added
mass, defined via the memory kernel, was negative, and the
backflow of neighbors tended to drag the particle in the direction
of its initial velocity, i.e. contributed negatively to the
friction.

The above mentioned theories yield the reliable results for the
different time domains: whereas the GCM allows to study the fluid
dynamics at the early and intermediate stage of the system
evolution, the MCT describes the long-time tails of the
corresponding TCF. Moreover, neither GCM nor MCT give possibility
to investigate a transition from the intermediate time domain to
the hydrodynamic stage. Recently, such an interesting problem has
been considered within an approach \cite{russians1,russians2} that
not only exploits some theoretical schemes, like those above
mentioned, but also uses the computer simulation results as the
necessary input data for the subsequent complex analysis of the
fluids dynamics (therefore, it is sometimes referred to as a
``combined'' method). In the cited papers, the continued fraction
method for the Laplace transforms of the VAFs and reasonable
approximations for the memory kernels were used with subsequent
analysis of the VAF peculiarities at the complex frequency plane.
It was shown that unlike the GCM case, characterized by a limited
number of the isolated poles, VAF shows the singularity manifold
forming branch cuts. The branch cuts were found to be separated
from the real axis by the well-defined ``gap''. The inverse value
of the gap width determines a duration of the transition period
from the short-time one-particle kinetics to the long-time
collective motion (hydrodynamics) with a typical power law
relaxation $t^{-3/2}$, what for VAF has been reported for the
first time and explained by Alder and Wainwright \cite{Ald70}.

Summarizing all the above mentioned, we would like to note that in
spite of the considerable achievements in exploration of the
fluids dynamics, there is still a lack of unified self-consistent
approaches, being able to describe the system behavior throughout
all the stages of its evolution. Thus, an attempt to create and
verify such an approach is a motivation of our study. A present
paper should be considered as a logical continuation of our very
recent article \cite{CMPour}, where the basic concept of the
effective summing up of the infinite continued fractions has been
formulated, and some relevant theoretical results for the VAFs
behavior have been obtained. However, in Ref.~\cite{CMPour} the
input parameters of the theory (some kinds of SCFs) had been
initially prescribed and did not rely upon the data of experiments
or computer simulations.

In the present paper, we take the next step, model\-ling the
dynamics of the realistic fluids. To this end, we perform computer
simulations of a number of the autocorrelation functions for the
liquid Ne at the temperature $T=295$ K and various densities in
the range $\rho=691\div 2190$ kg/m$^3$. The results obtained
within our theoretical scheme are compared with the MD data as
well as with those derived by the MA and predicted by the
``combined'' theories \cite{russians1,russians2}.

A structure of the paper is as follows. In Sec.~\ref{secII}, we
briefly recall the method of continued fractions for the VAF
presentation and calculate the kinetic kernels of the highest
order at a physically reasonable assumption adopted in
Ref.~\cite{CMPour}. In Sec.~\ref{secIII}, we evaluate the spectral
functions both by the effective summation of the continued
fraction and in the finite modes approximation. The MD simulation
of the VAFs as well as the autocorrelation functions, defined by
the higher derivatives of the velocity up to the 4-th order, with
a subsequent analysis of the obtained results is a subject of
Sec.~\ref{secIV}. In Sec.~\ref{secV}, we study a time behavior of
the VAFs, obtained within different theoretical schemes, and
compare the results with the computer simulation data. Finally, we
draw the conclusions in Sec.~\ref{secVI}.

\section{Continued fraction representation for velocity autocorrelation function\label{secII}}

To begin with the study of the fluids dynamics, let us consider
the normalized VAF \bea\label{VAF}F_v(t)=\langle \textbf
v_i\exp(-i L_N t)\textbf v_i\rangle/\langle \textbf v_i\textbf
v_i\rangle,\eea where $i L_N$ means the Liouville operator. For
the system of structureless particles interacting via the central
force potential $U(|\textbf r_j-\textbf r_l|)$, it has a quite
simple form,
$$
i
L_N=\sum\limits_{j=1}^N\left(\frac{\textbf{p}_j}{m_j}\frac{\partial}{\partial\textbf
r}_j-\sum\limits_{j\ne j=1}^N\frac{\partial U(|\textbf r_j-\textbf
r_l|)}{\partial\textbf r_j}\frac{\partial}{\partial\textbf
p_j}\right),
$$
with positions $\textbf r_j$ and momenta $\textbf p_j$ of
particles. The Laplace-transformed VAF $\tilde
F_v(z)=\int_0^{\infty}\exp(-z t) F_v(t) dt$ obeys the equation
\bea\label{tFv} \tilde F_v(z)=\frac{1}{z+\tilde{\phi}_1(z)}, \eea
where $\tilde{\phi}_1(z)$ are the Laplace-components of the lowest
order memory kernel, which contains all the information about the
dissipation processes in fluid. What is important for the
statement of problem: Eq.~(\ref{tFv}) is, in fact, an identity
like the well-known Ornstein-Zernike equations in the equilibrium
theory of the fluids \cite{OZ}. Thus, any specific dynamics of the
kinetic kernels like long-time tails (if any noticeable) would
manifest itself in the corresponding VAFs \cite{MHforMemKer}.

An explicit expression for $\tilde F_v(z)$ can be obtained from
the higher order memory kernels $\tilde{\phi}_j(z)$, which satisfy
a recurrence relation \cite{Boo} \bea\label{recurrV}
\tilde{\phi}_{j-1}(z)=\frac{\Gamma_{j-1}}{z+\tilde{\phi}_{j}(z)},\eea
where $\Gamma_{j-1}$ is the relevant SCF. Equations
(\ref{tFv})-(\ref{recurrV}) can be also rewritten in time
representation, \bea\label{Eqns-time} \nonumber &&\frac{\partial
F_v(t)}{\partial
t}+\int\limits_0^t dt'\phi_1(t-t') F_v(t')=0,\\
\nonumber
&&\ldots\\
 &&\frac{\partial \phi_{j-1}(t)}{\partial t}+\int\limits_0^t
dt'\phi_j(t-t') \phi_{j-1}(t')=0,
 \eea
 where $\phi_j(t=0)=\Gamma_j.$ This is a hierarchy of the generalized Langevin equations, in
which the kinetic kernels $\phi_j(t)$ are some TCFs of the special
kind, defined by the ``fluctuating forces'' \cite{Mori}.

Using Eqs.~(\ref{tFv})-(\ref{recurrV}), it is possible to present
the equation for the Laplace transform $\tilde F_v(z)$ as an
infinite continued fraction \bea\label{VAFfrac} \tilde F_v(z)=
\displaystyle\displaystyle\frac{1}{z+\displaystyle\frac{\Gamma_1}{z+\displaystyle\frac{\Gamma_2}{z+\displaystyle\frac{\Gamma_3}{z+\cdots}}}}.
\eea All the SCFs $\Gamma_j$ entering Eq.~(\ref{VAFfrac}) can be
calculated according to the general definition \cite{Boo}. The
lowest order SCFs can be presented as follows: \bea\label{Gamma1}
\Gamma_1=\frac{1}{m^2}\frac{\langle\mathbf F_i \mathbf
F_i\rangle}{\langle\mathbf v_i \mathbf v_i\rangle}=
\frac{\beta}{3m}\langle\mathbf F_i \mathbf F_i\rangle,\eea
\bea\label{Gamma2}&&\Gamma_2=\frac{\langle{iL_N\mathbf F_i
iL_N\mathbf F_i}\rangle}{\langle\mathbf F_i \mathbf
F_i\rangle}-\Gamma_1. \eea In Eqs.~(\ref{Gamma1})-(\ref{Gamma2}),
$\mathbf F_i$ means the force acting on the $i$-th particle,
$\mathbf v_i$ denotes its velocity and the brackets denote
ensemble averaging. The ``force--force'' SCF (\ref{Gamma1}) can be
evaluated using a pair correlation function of the fluid at a
particular thermodynamic point. The SCFs $\Gamma_j$ with $j\ge 2$
would require knowledge of even higher order correlation
functions, rendering such a problem completely unmanageable.
Therefore, the reliable results for the higher order SCFs as well
as for the VAFs themselves have to be obtained only by computer
simulations (see Sec.~IV for the details).

The infinite continued fraction (\ref{VAFfrac}) can be rewritten
\cite{Bafile1} as a weighted sum \bea\label{sumZ} \tilde
F_v(z)=\sum\limits_{\alpha=1}^{\infty}
\frac{I_{\alpha}}{z-z_{\alpha}}, \eea where $I_{\alpha}$ denote
the amplitudes of the particular $\alpha$-th mode of the fluid,
whereas $z_\alpha$ mean the mode frequencies. Since evaluation of
the high order SCFs $\Gamma_j$ does not seem a realistic problem
from the viewpoint of structural analysis of the fluid or requires
time consuming computer simulations, for practical purposes one
has to truncate the series (\ref{sumZ}), retaining only a limited
number $M$ of terms \cite{Bafile2}. In the general case, the
issue, at which hierarchy level such a truncation should be
performed, has to be solved separately for each particular system
and for each particular thermodynamic point
\cite{Bafile2,Bafile3}.

A simple truncation of the series (\ref{sumZ}), even at large
number of terms, does not allow to describe the system dynamics
exactly, especially at long times \cite{MCT1,MCT2}. On the other
hand, from a strictly mathematical viewpoint, there is no other
reason for the power law behavior of the VAFs at large $t$, but a
contribution of the infinite number of the terms in (\ref{sumZ}).
Therefore, it would be challenging to propose a self-consistent
theory that could reliably describe the fluids dynamics within the
whole time domain with: i) taking advantages of the GCM or MA
approaches, and ii) its application for the long times at the
expense of the properly designed kinetic kernels.

To this end, let us make an important remark. The Markovian
approximation of the lowest order kinetic kernel is known
\cite{mryglod98} to determine the relaxation time $\tau_1$ of the
corresponding hydrodynamic excitation by a simple relation
$\tilde{\phi}_1(0)=\tau_1^{-1}$. Further natural step would be an
introduction (in a similar manner) of the higher order relaxation
times, $\tau_j$, with $j\ge 2$. It is known from the theory of
continued fractions \cite{Campos} as well as from the analysis of
fluid dynamics \cite{our_assump2} that $\tau_j$ oscillate around a
certain value $\tau^*$ depending on the level of description, and
approach this asymptotics from below (odd order relaxation times)
or above (even order relaxation times). Moreover, the difference
between $\tau_{2j}$ and $\tau_{2j+1}$ decreases rapidly with
increasing of the hierarchy level $j$. A key point of our approach
is a physically grounded assumption that the relaxation times of
the high order kinetic kernels become comparable with each other
and, starting from the order $s$, tend to a certain fixed value,
which has a purely kinetic origin. In this context, our assumption
is in a close agreement with the results
\cite{our_assump2,PRE_polar} for characteristic times of decay of
the autocorrelation functions of simple and polar fluids, obtained
by an approximated summation of the continued fractions.

The above formulated statement can be formalized in the following
way. Let us suppose that the memory kernels $\tilde{\phi}_j(z)$
tend to a certain (limiting) function with the increase of $j$. If
we suppose this convergence to occur for two kinetic kernels of
the neighboring orders, starting from large enough $j=s$,
\bea\label{approxS} \tilde{\phi}_s(z)\approx\tilde{\phi}_{s+1}(z),
\eea then the recurrence relation (\ref{recurrV}) converts into a
bilinear form, which gives us immediately the explicit expression
for the highest order memory kernel \bea\label{phiS_result}
\tilde{\phi}_s(z)=-\frac{z}{2}+\sqrt{\frac{z^2}{4}+\Gamma_s}. \eea
The relation (\ref{approxS}) is, in fact, the $s$-th order
approximation for the memory kernels at the chosen ansatz. In
contrast to the Markovian approximation, \bea\label{MA}
\tilde{\phi}_s(z)\approx\tilde{\phi}_{s}(0), \eea when the real
part of the kinetic kernel $\tilde{\phi}_s(z=i\omega)$ is a
constant in the whole frequency domain, whereas the imaginary one
equals to zero, the expression (\ref{phiS_result}) takes into
account the frequency dispersion of the memory function, which
will be discussed a bit later. We shall refer to the relation
(\ref{MA}) as the Markovian approximation at the $s$-th hierarchy
level (MAs), introducing an extra symbol $s$ in the above
mentioned abbreviation. In the subsequent Sections, we consider
the kinetic kernels for a particular number of $s=2\div 5$, or,
what is the same, use the approximations (MA2)$\div$(MA5).
Hereafter, the index $s$, which denotes the chosen approximation
level and is related to the highest order kinetic kernel, should
not be confused with the ``current'' indexes $j\le s$.

In particular, it would be instructive to compare the result for
the VAF in the MA2 approximation with that obtained within the GCM
theory \cite{GCM4}. To this end, let us use the recurrence
relation (\ref{recurrV}) along with the MA for the highest order
memory kernel: \bea\label{phiS_Mark} \tilde{\phi}_{s-1}(z)\approx
\frac{\Gamma_{s-1}}{z+\tilde{\phi}_s(0)}
=\frac{\Gamma_{s-1}}{z+\sqrt{\Gamma_s}}.\eea Taking $s=2$ in
Eq.~(\ref{phiS_Mark}) and relating the SCFs $\Gamma_1$, $\Gamma_2$
to the self-diffusion coefficient $D=1/3\int_0^{\infty}dt \langle
\textbf{v}_i(t)\textbf{v}_i(0)\rangle$ by the equation
$\displaystyle\frac{\sqrt{\Gamma_2}}{\Gamma_1}=\displaystyle\frac{m
D}{k_B T}$, we can reproduce the two-modes expression (15) of
Ref.~\cite{GCM4} for the VAF of a simple fluid. In the higher
order approximations, this result can be generalized as
\bea\label{DSmany} \frac{m D}{k_B T}=\left\{
\begin{array}{ccc}
\displaystyle\frac{\Gamma_2\cdot\Gamma_4\cdots\Gamma_{2k}}{\Gamma_1\cdot\Gamma_3\cdots\Gamma_{2k-1}}&\displaystyle\frac{1}{\sqrt{\Gamma_{2k+1}}},&
s=2k+1
\\[3ex]
\displaystyle\frac{\Gamma_2\cdot\Gamma_4\cdots\Gamma_{2k-2}}{\Gamma_1\cdot\Gamma_3\cdots\Gamma_{2k-1}}&\sqrt{\Gamma_{2k}},&
s=2k,
\end{array}\right.
\eea
allowing one to compare the results obtained within the MAs with
those calculated in the framework of the GCM $s$-modes
approximation. However, as we have already mentioned, this
equivalence with the GCM is mainly from a mathematical viewpoint,
since in the GCM the self-diffusion coefficient is determined via
a zero-time moment of the VAF, being calculated by the computer
simulations. In the MAs, we use the SCFs $\Gamma_j$ as the input
parameters (which also should be evaluated in computer
simulations) and do not concern in the correlation time
$\tau_{dif}=m D/k_B T$, related to the self-diffusion.

It is worthy to note that the real part,
$\mbox{Re}[\tilde{\phi}_{s-1}(i\omega)]$, of the Lorenz-type
kinetic kernel (\ref{phiS_Mark}) decays as $\omega^{-2}$ in the
high frequency domain, whereas the imaginary part decays as
$\omega^{-1}$.  In time representation, it corresponds to the
exponential relaxation. In its turn, the real part of the highest
order kinetic kernel $\tilde{\phi}_s(i\omega)$, defined by
Eq.~(\ref{phiS_result}), sharply vanishes at the cut-off frequency
$\omega_c=2\sqrt{\Gamma_s}$, while its imaginary part after the
lapse of the linear growth decays as $\omega^{-1}$ at large
frequencies. It can be easily shown using the recurrence relation
(\ref{recurrV}) that the same is true for the highest but one
order memory kernel, \bea\label{phiS-1}
\tilde{\phi}_{s-1}(i\omega)=\frac{\Gamma_{s-1}}{\Gamma_{s}}\tilde{\phi}_{s}(i\omega).
\eea Moreover, these results look even more interesting, if one
performs the inverse Laplace transformation to
(\ref{phiS_result}), \bea\label{phiS_t}
\phi_s(t)\equiv\frac{1}{2\pi i}\lim\limits_{\epsilon\to
0}\!\!\int\limits_{\epsilon-i\infty}^{\epsilon+i\infty}\!\!\! e^{z
t}\tilde{\phi}_s(z)dz
=\sqrt{\Gamma_s}\,\,\frac{J_1(2\sqrt{\Gamma_s}t)}{t}, \eea where
the Bessel function of the first order $J_1$ appears. It means
that the $s$-th and the $(s-1)$-th order me\-mo\-ry kernels decay
non-monotonically at long times as $t^{-3/2}$, since the Bessel
function $J_1$ has the well known asymptote $J_1(2\sqrt{\Gamma_s}
t\gg 1)=\cos(2\sqrt{\Gamma_s}(t-3\pi/4))/\sqrt{\pi t}.$ Thus,
their dynamics completely differs from the time behavior of the
highest order memory functions in the MA, which are
$\delta$-correlated in time, and from the $(s-1)$-th order kinetic
kernels in the MA, which are found to have an exponential
relaxation, see Eq.~(\ref{phiS_Mark}).

Eqs.~(\ref{phiS_result}) and (\ref{phiS_t}), obtained in the
framework of the rigorous approach using just one physically
reasonable assumption (\ref{approxS}), are the cornerstones for
our further study of the VAFs dynamics. From the mathematical
point of view, we have done nothing but an effective summation of
the infinite continued fraction (\ref{VAFfrac}), in which all the
parameters $\Gamma_j$ are equal to each other starting from the
certain value $j=s$. Thus, in fact, we deal with the problem with
a limited number $s$ of the parameters like it is within the GCM
(MA) approaches. In the next Section we consider the spectral
functions with various values of $s$, at which the condition
(\ref{approxS}) are supposed to hold true.

\section{Spectral functions\label{secIII}}

Let's consider the spectral function (SF) $\tilde
F_v(\omega)=\mbox{Re}\,[\tilde F_v(z=i\omega)]$, defined as a real
part of the corresponding continued fraction (\ref{VAFfrac}) taken
at the imaginary frequency. In this Section, we replace the label
$v$ in the expressions for SFs by the subscript $\alpha$, which
denotes the chosen level of approximation (see below for details).

At the beginning, let us adopt the approximation of the second
hierarchy level. In such a case, it is straightforward to evaluate
the SF $\tilde F_2(\omega)$, \bea\label{spectr2} \tilde
F_2(\omega)=\frac{\Gamma_1\sqrt{\Gamma_2-\omega^2/4}}{(\Gamma_2-\Gamma_1)\omega^2+\Gamma^2_1}.
\eea It is also instructive to compare the result (\ref{spectr2})
with the expression \bea\label{spectrM2} \tilde
F_{M2}(\omega)=\frac{\Gamma_1\sqrt{\Gamma_2}}{\omega^4+(\Gamma_2-2\Gamma_1)\omega^2+\Gamma_1^2}
\eea for the SF, obtained by a truncation of the continued
fraction at the level $s=2$. Hereafter, we use the extra subscript
$M$ to denote the above mentioned MA (or modes) approximation.

It is useful to compare the expressions
(\ref{spectr2})-(\ref{spectrM2}) for the corresponding SFs. First
of all, the former spectral function vanishes at the cut-off
frequency $\omega_c=2\sqrt{\Gamma_2}$, while the latter tends to
zero asymptotically as $1/\omega^4$.

Secondly, both SFs are even functions of frequency and behave as
$\lim_{\omega\to 0}d\tilde F_{\alpha}(\omega)/d\omega=0$, here
$\alpha$ denotes 2 or M2. Thirdly, in the domain of frequencies
$\omega\ll\omega_c$ and at $\Gamma_1\ll\Gamma_2$, the relation
$\tilde F_2(\omega)\approx\tilde F_{M2}(\omega)$ holds true. It
means that at a certain relation between the parameters
$\Gamma_j$, the results, obtained within our approximation and the
GCM (MA) formalisms, are close to each other (we will touch upon
this issue in more detail in Sec.~V, when analyzing the time
behavior of the corresponding VAFs).

At last but not least, in the vicinity $\varepsilon$ of the
cut-off frequency $\omega_c$ the SF (\ref{spectr2}) behaves as
\bea\label{sqrtEps} \tilde
F_2(\omega_c-\varepsilon)\approx\frac{\Gamma_1\Gamma_2^{1/4}}{(\Gamma_1-2\Gamma_2)^2}\sqrt{\varepsilon}.
\eea The square root dependence (\ref{sqrtEps}) resembles the
well-known result from the MCT \cite{MCT3,long_tail1}. However, in
the MCT approach a non-analytical dependence $\sqrt\omega$ of the
SFs, generating the long-time tails $\sim t^{-3/2}$ of the VAFs,
occurs at the zero frequency domain rather than at $\omega_c$.

A similar structure of the SF can be obtained at the approximation
of the third hierarchy level. In this case, the spectral function
$\tilde F_3(\omega)$, obtained by summing up the continued
fraction for the 3-rd order memory kernel, can be presented as
follows: \bea\label{spectr3} \tilde
F_3(\omega)=\frac{\Gamma_1\Gamma_2\sqrt{\Gamma_3-\omega^2/4}}{(\Gamma_3-\Gamma_2)\omega^4+
\left(\Gamma_2(\Gamma_1+\Gamma_2)-2\Gamma_1\Gamma_3
\right)\omega^2+\Gamma_1^2\Gamma_3}, \eea while its counterpart in
the modes approximation looks as
\bea\label{spectrM3} \tilde F_{M3}(\omega)=
\frac{\Gamma_1\Gamma_2\sqrt{\Gamma_3}} {\omega^6-[
2(\Gamma_1+\Gamma_2)-\Gamma_3]\omega^4+[
(\Gamma_1+\Gamma_2)^2-2\Gamma_1\Gamma_3
]\omega^2+\Gamma_1^2\Gamma_3}. \eea

Once again, like in the the cases
(\ref{spectr2})-(\ref{spectrM2}), two SFs are close to each other,
$\tilde F_3(\omega)\approx\tilde F_{M3}(\omega)$ in the domain of
frequencies $\omega\ll\omega_c$ and at
$\Gamma_3\gg\mbox{max}\{\Gamma_1,\Gamma_2\}$.

Inspecting Eq.~(\ref{spectr2}) and taking into account the sum
rule \bea\label{sumRule0}
\frac{1}{2\pi}\int\limits_{-\infty}^{\infty}\tilde
F_s(\omega)d\omega\equiv\frac{1}{2\pi}\int\limits_{-\omega_c}^{\omega_c}\tilde
F_s(\omega)d\omega=1, \eea (which, in fact, denotes the initial
value of the VAF being equal to the unity), one can observe by a
direct integration that there is a minimal value
$\Gamma_2^{(min)}=\Gamma_1/2$, below which the condition
Eq.~(\ref{sumRule0}) is no longer valid.

In a similar way, the value
$\Gamma_3^{(min)}=(\Gamma_1+2\Gamma_2)/4$ can be defined for the
fixed (lower order) SCFs $\Gamma_1$, $\Gamma_2$ in
Eq.~(\ref{spectr3}), and the sum rule (\ref{sumRule0}) is
satisfied for $\Gamma_3\ge\Gamma_3^{(min)}$. In this context, our
results differ from those (see Eqs.~(\ref{spectrM2}),
(\ref{spectrM3})), obtained within the modes approximation: in the
latter case, no relation between various $\Gamma_i$ is needed for
the sum rule (\ref{sumRule0}) to be satisfied.

We have to emphasize that so far the minimal values
$\Gamma_s^{(min)}$ appear in a strictly ``mathematical'' way due
to the requirement (\ref{sumRule0}). In a recent paper
Ref.~\cite{CMPour}, we have
 attributed more physical meaning for
$\Gamma_s^{(min)}$. This parameter was shown to be the threshold
value, which defines a crossover from the power law $\sim
t^{-3/2}$ to the slower relaxation $\sim t^{-1/2}$ of the VAF,
when the fluid starts to behave like a typical 1D-system
\cite{indians}. However, computer simulation data for the fluids
show (see Sec.~IV) that the highest order SCFs  are well above
their minimally allowed values $\Gamma_s^{(min)}$. Thus, a
possible solid-like scenario of the fluid dynamics, which has been
discussed in Ref.~\cite{CMPour}, is not realized in our case.

It is straightforward to obtain the SFs similar to
(\ref{spectr2})-(\ref{spectrM2}), (\ref{spectr3})-(\ref{spectrM3})
at any value of the hierarchy order $s$. Like in the above
presented case, the highest order SCF $\Gamma_s$ would determine
the corresponding cut-off frequency $\omega_c=2\sqrt\Gamma_s$. The
frequency expansion of the obtained SFs in the vicinity of
$\omega_c$ would also start from the square root term (see
Eq.~(\ref{sqrtEps})) with the renormalized coefficient due to
taking into account of the higher order $\Gamma_s$.

The above mentioned functions can be the starting points that
allow us to study a time behavior of the corresponding VAFs, using
the values for $\Gamma_j$, $j=\{1,\ldots,s\}$, taken from the MD
simulations. The computer simulation data as well as the results
of the theoretical prediction are presented in the next two
Sections.

\section{Time behavior of the autocorrelation functions: computer simulations\label{secIV}}

In our recent paper \cite{CMPour} we have calculated the VAFs at
various values of $\Gamma_j$, regardless of their relations to the
fluid at a particular thermodynamic point. We have presented
different regimes of the fluids dynamics (when the system behaves
in a solid-like, liquid-like or gas-like manner), modelled in the
framework of such an approach. Being aware that some (initially
prescribed) combinations of $\Gamma_j$ can be unrepresentative for
the real fluids, in this Section we study the dynamics of Ne at
some particular thermodynamic points.

\begin{figure*}[htb]
\centerline{\includegraphics[height=0.168\textheight,angle=0]{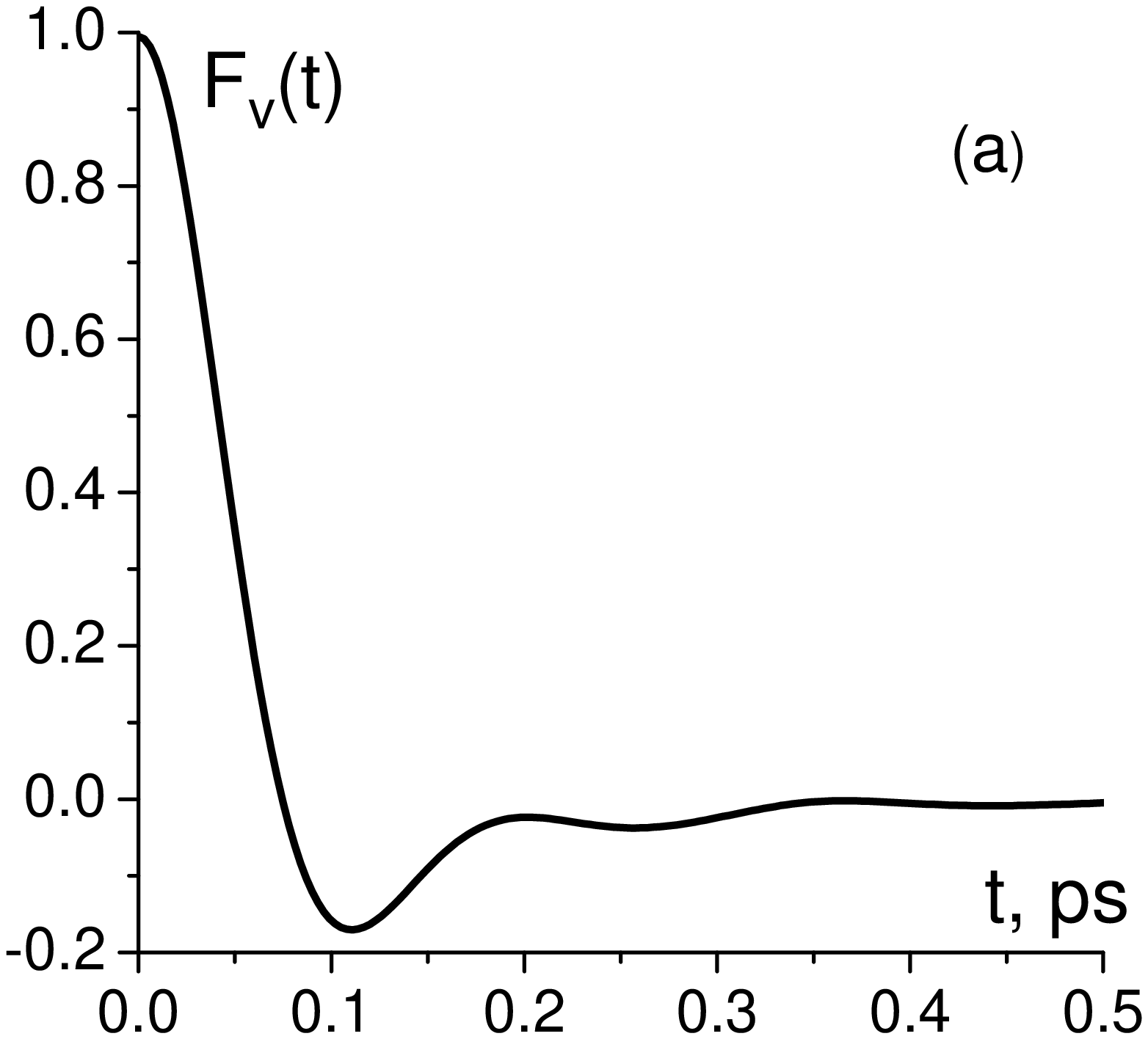}
 \includegraphics[height=0.168\textheight,angle=0]{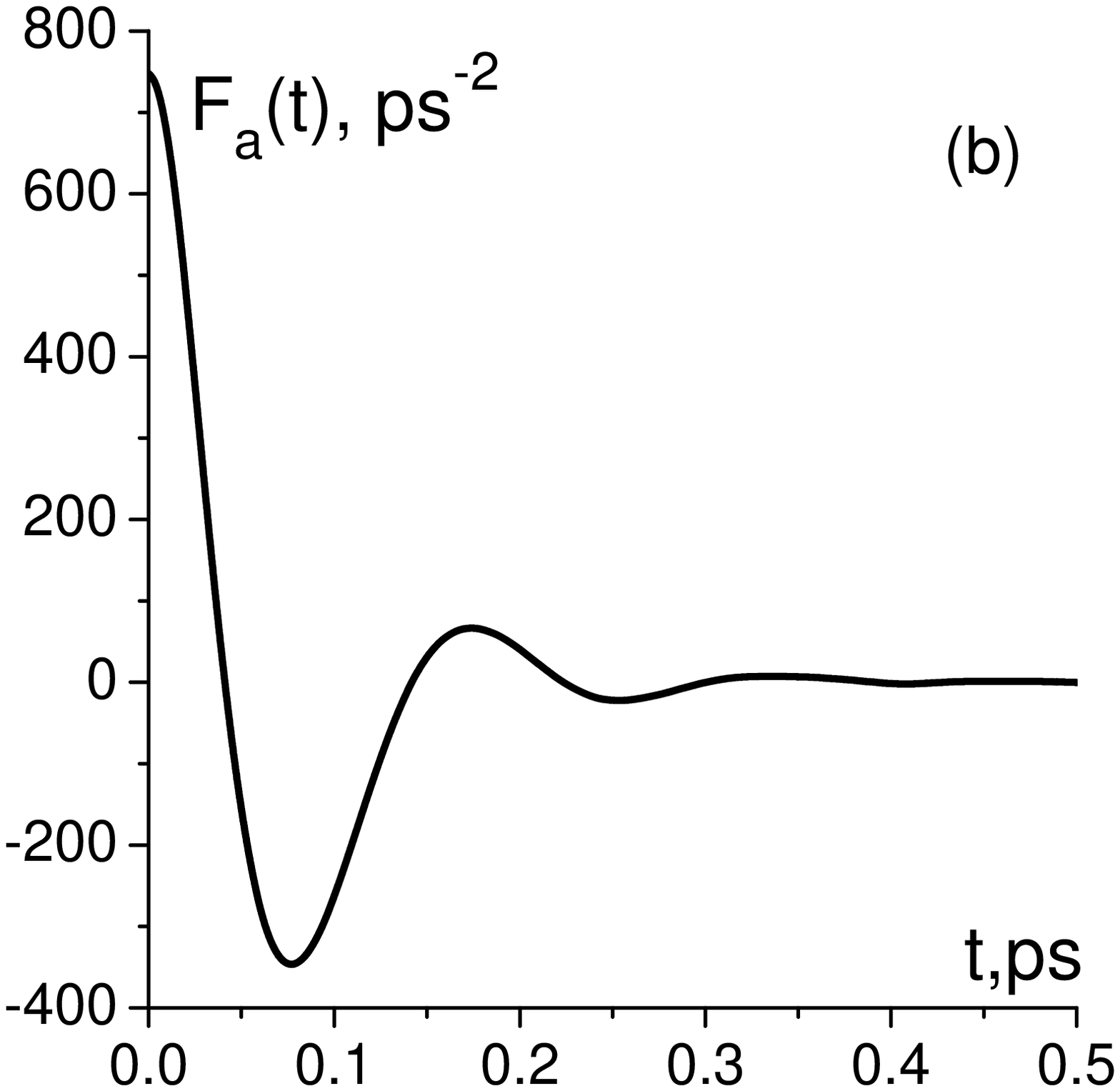}
  \includegraphics[height=0.168\textheight,angle=0]{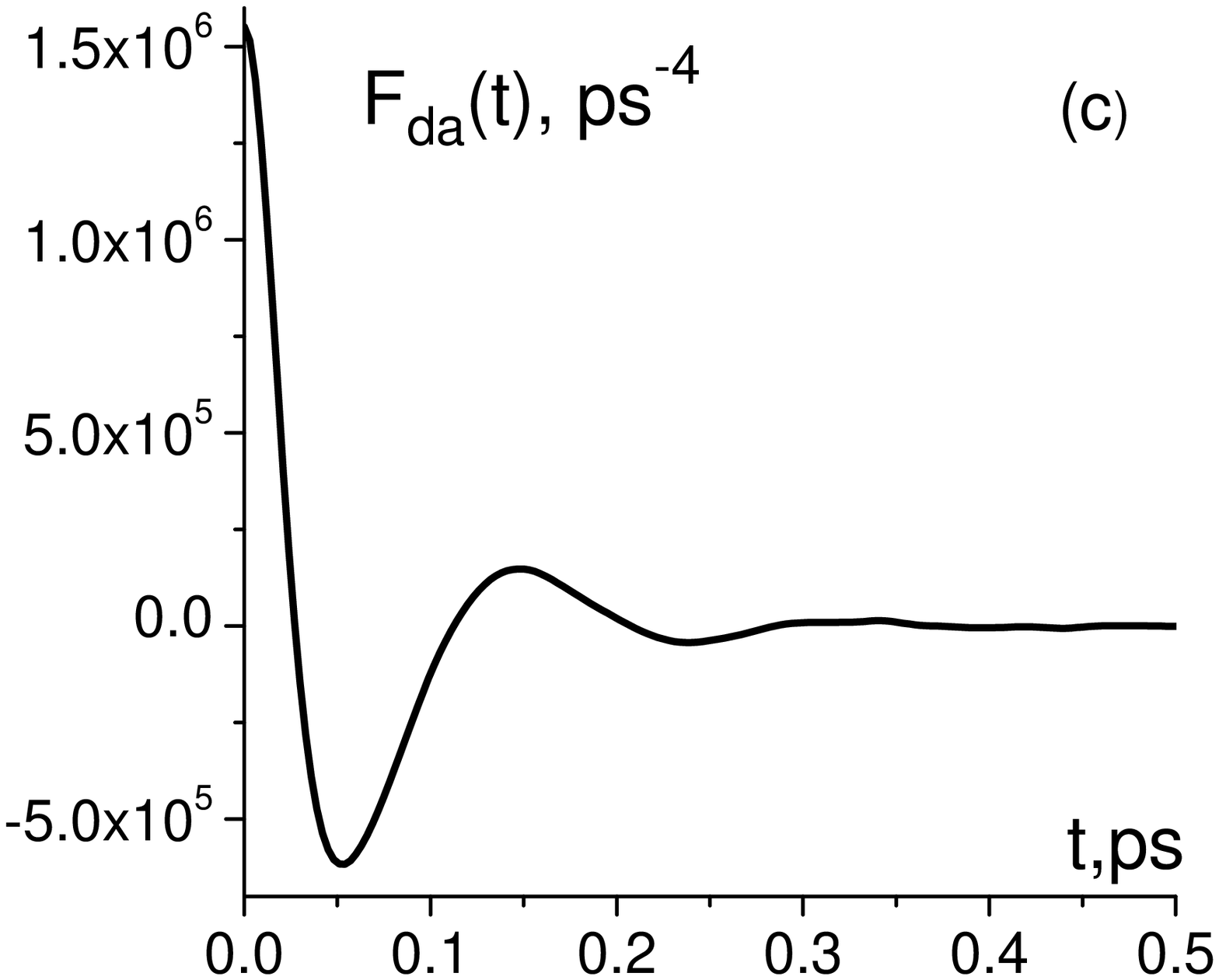}
   \includegraphics[height=0.168\textheight,angle=0]{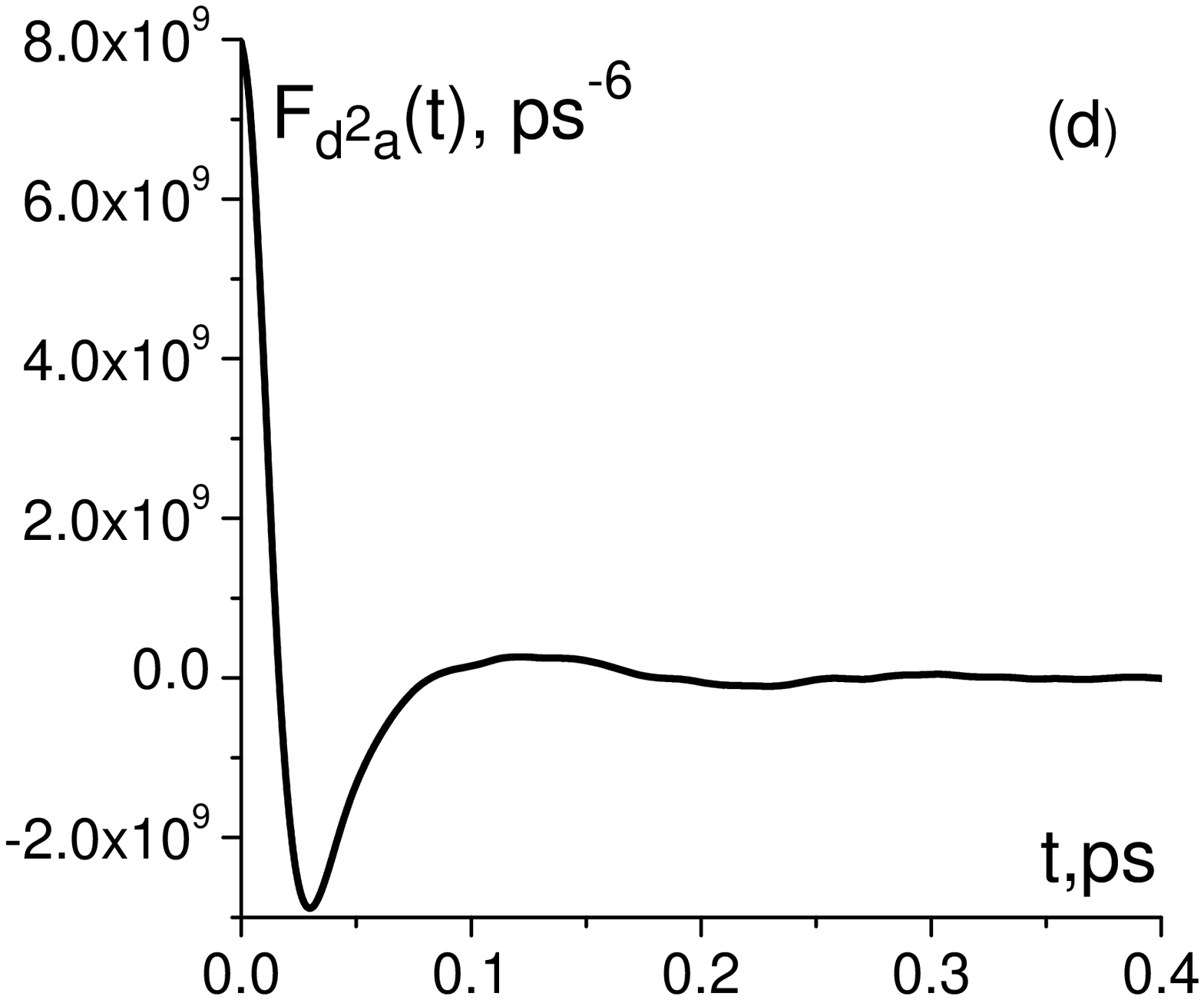}}
   \vspace{0.2cm}
   \centerline{\hspace{0.3cm}\includegraphics[height=0.181\textheight,angle=0]{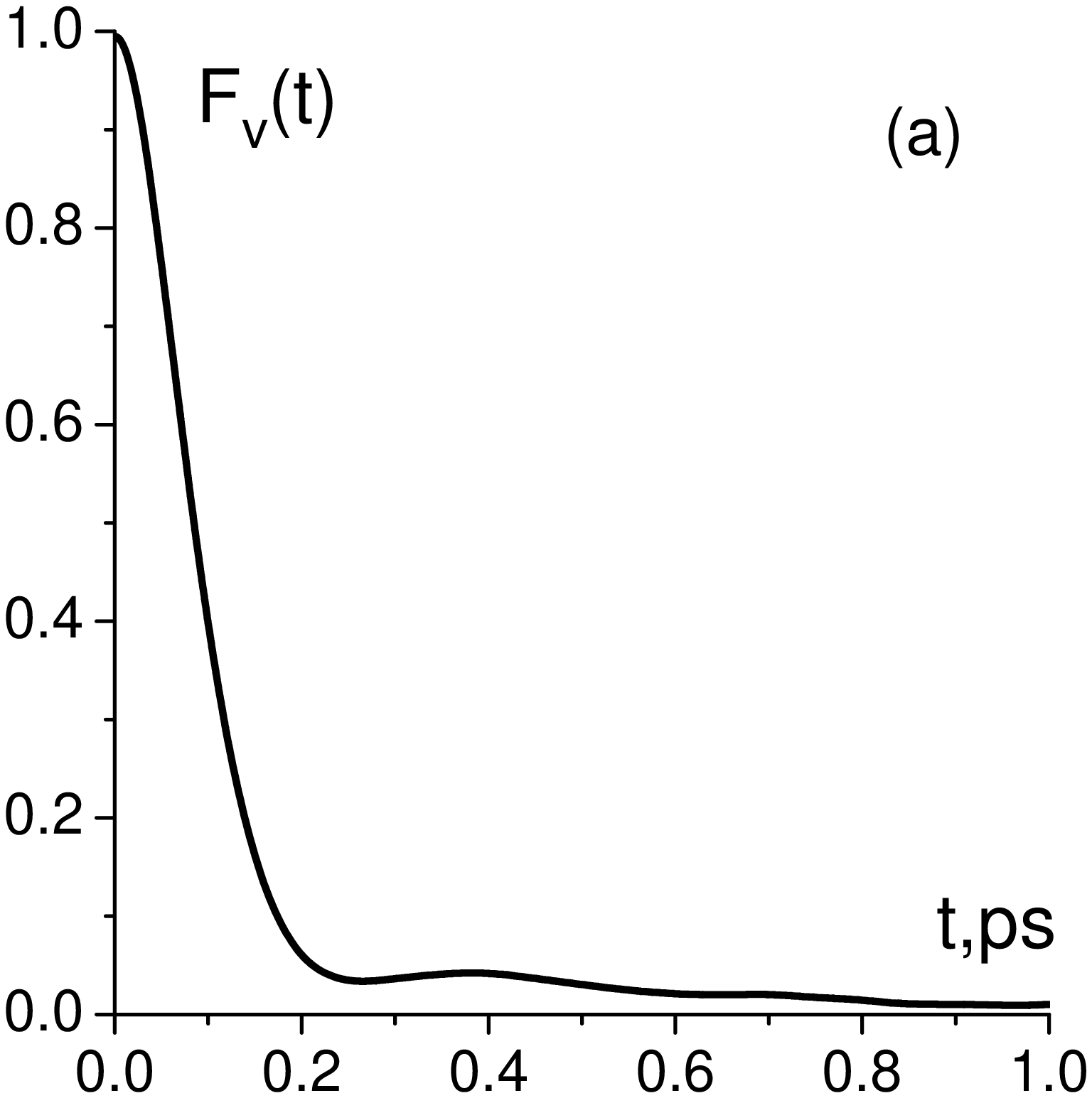}
 \includegraphics[height=0.177\textheight,angle=0]{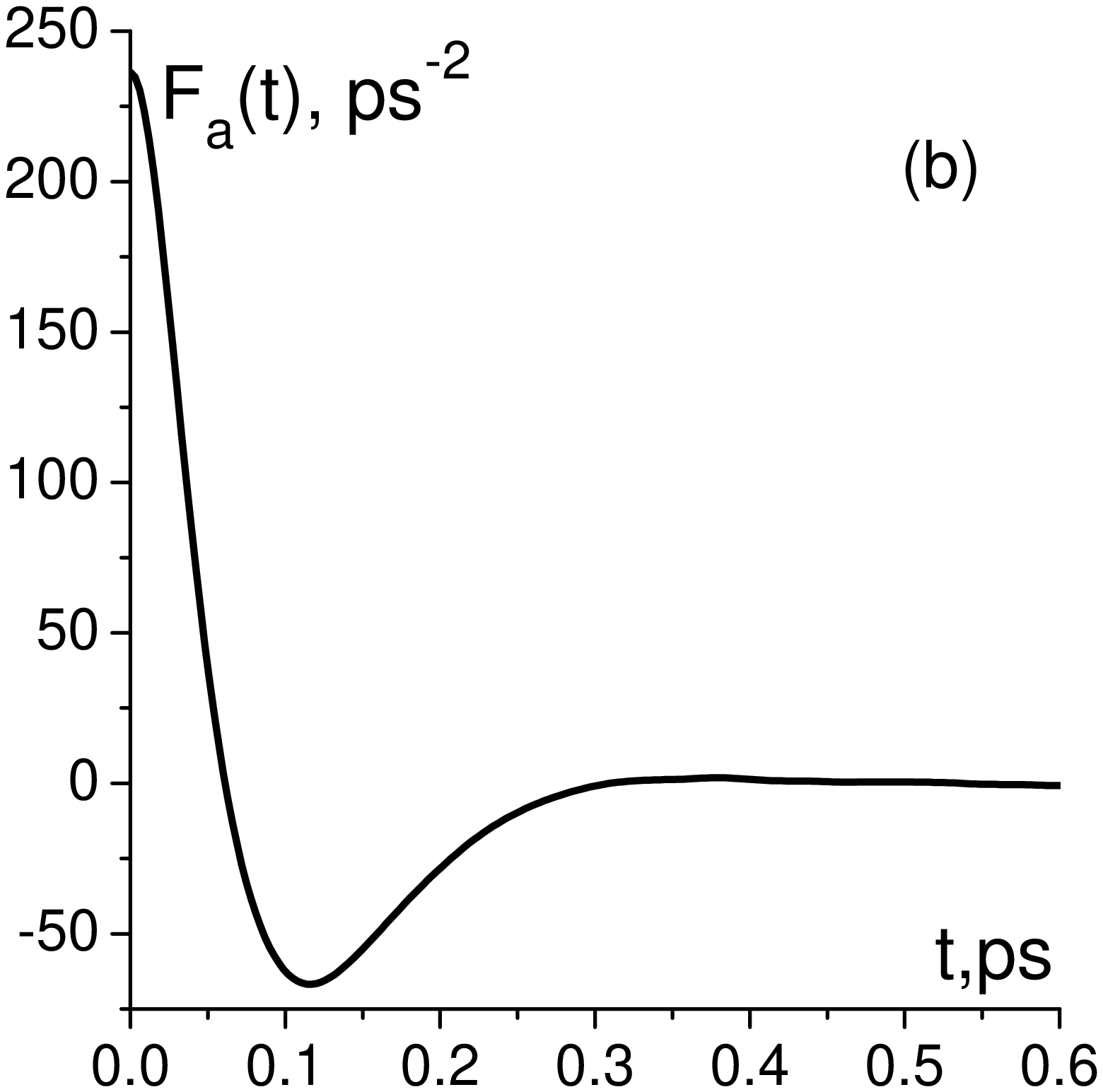}
  \includegraphics[height=0.177\textheight,angle=0]{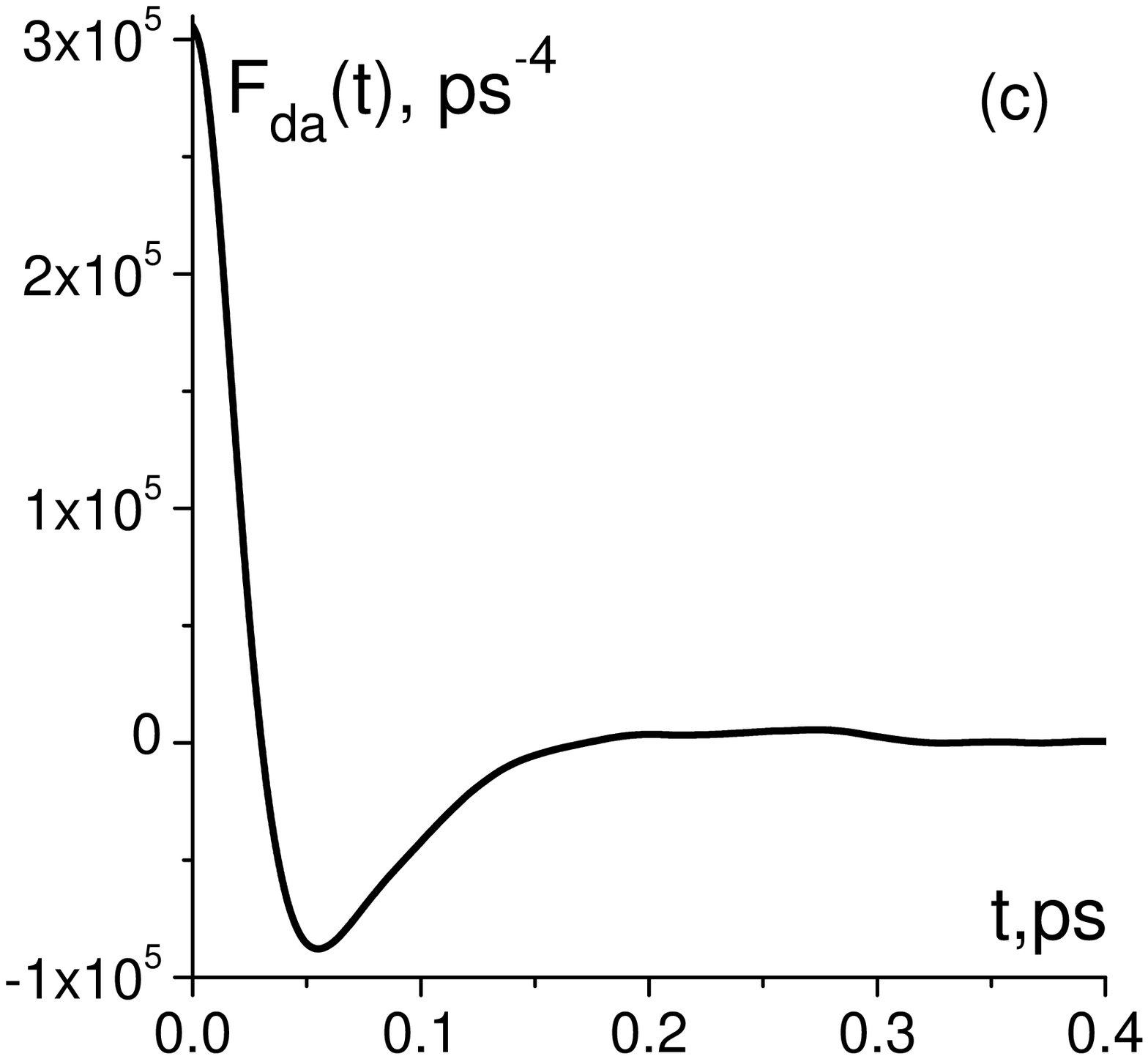}
   \includegraphics[height=0.177\textheight,angle=0]{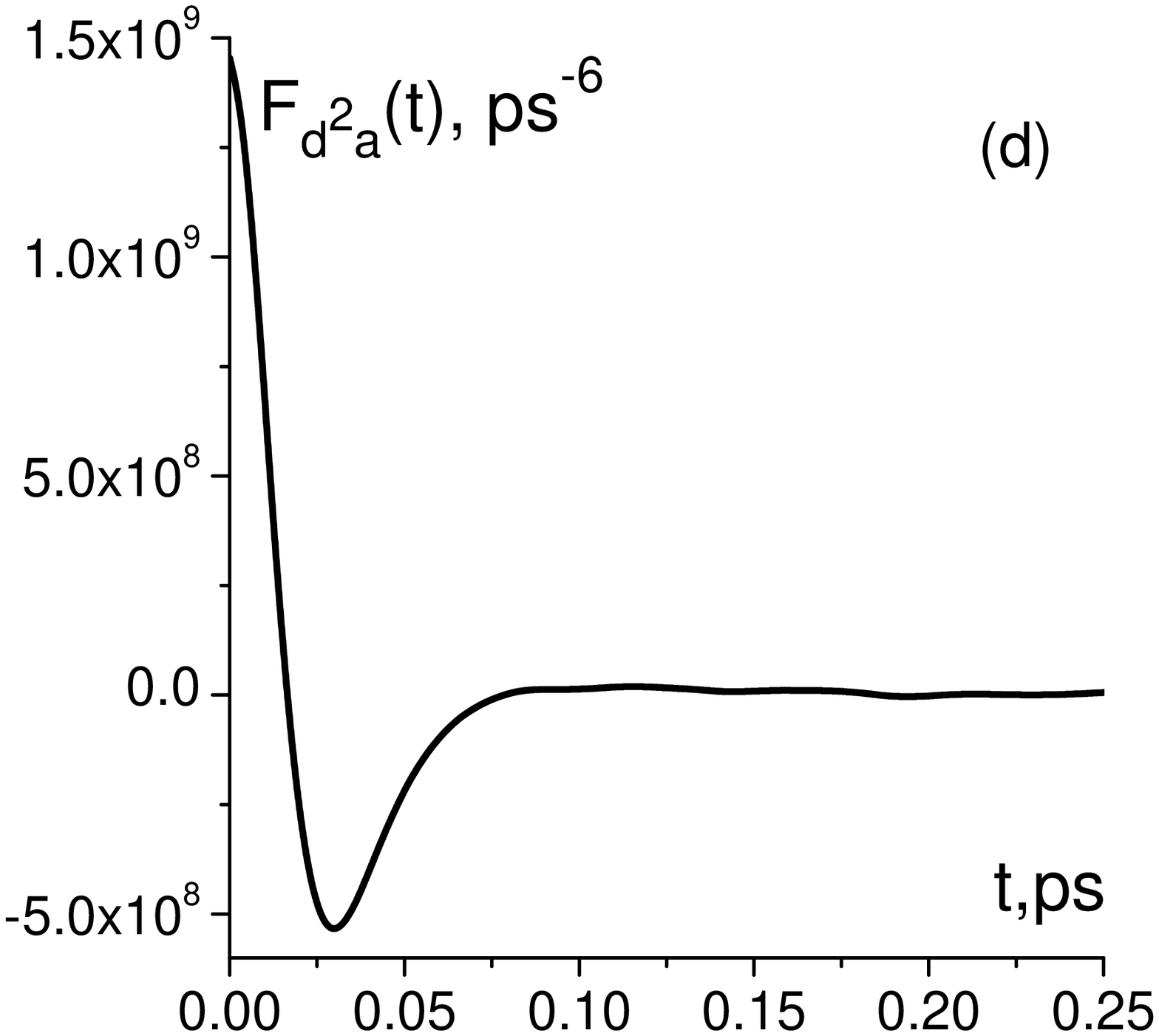}}
   \vspace{0.2cm}
   \centerline{\includegraphics[height=0.179\textheight,angle=0]{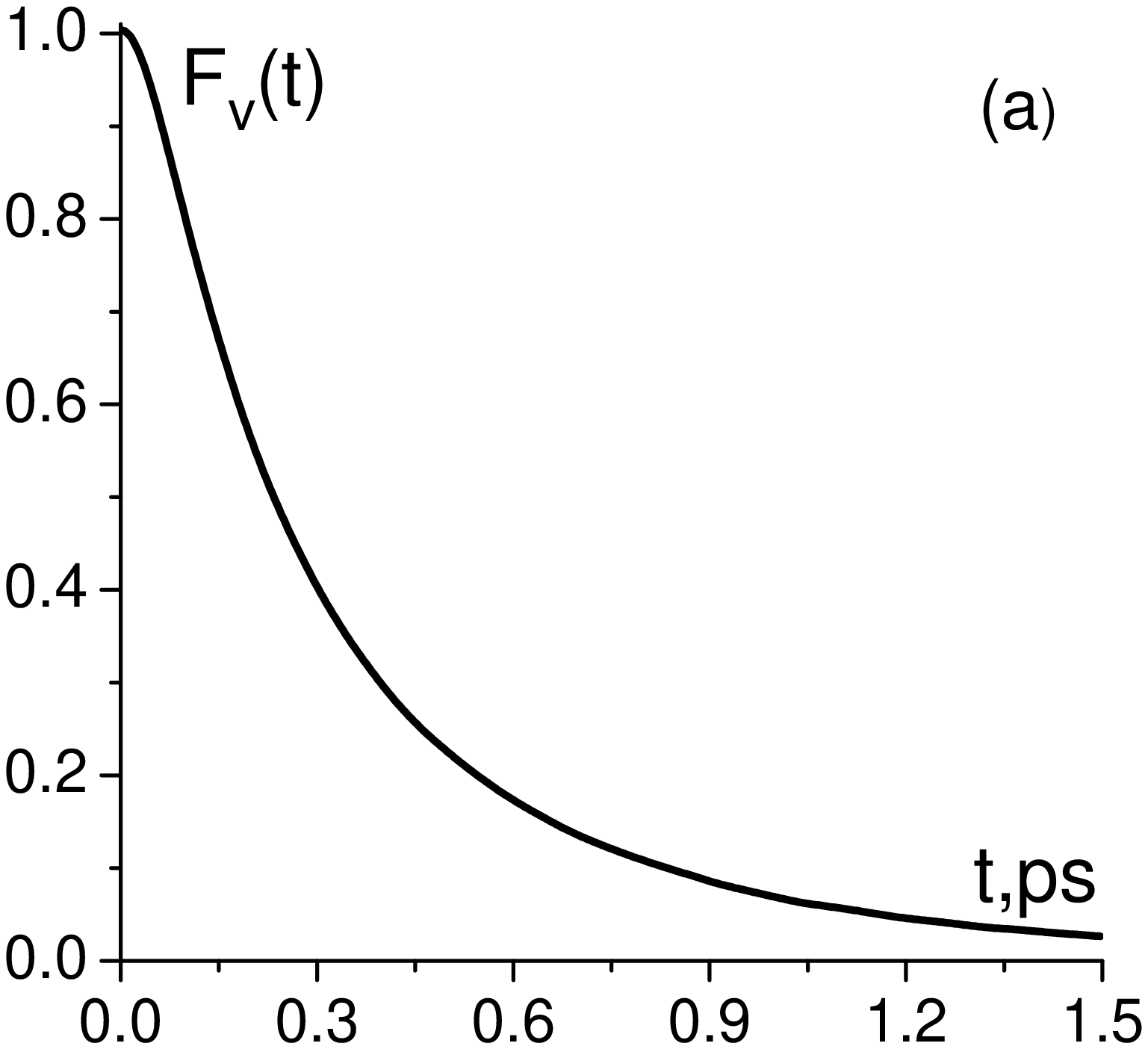}
 \includegraphics[height=0.179\textheight,angle=0]{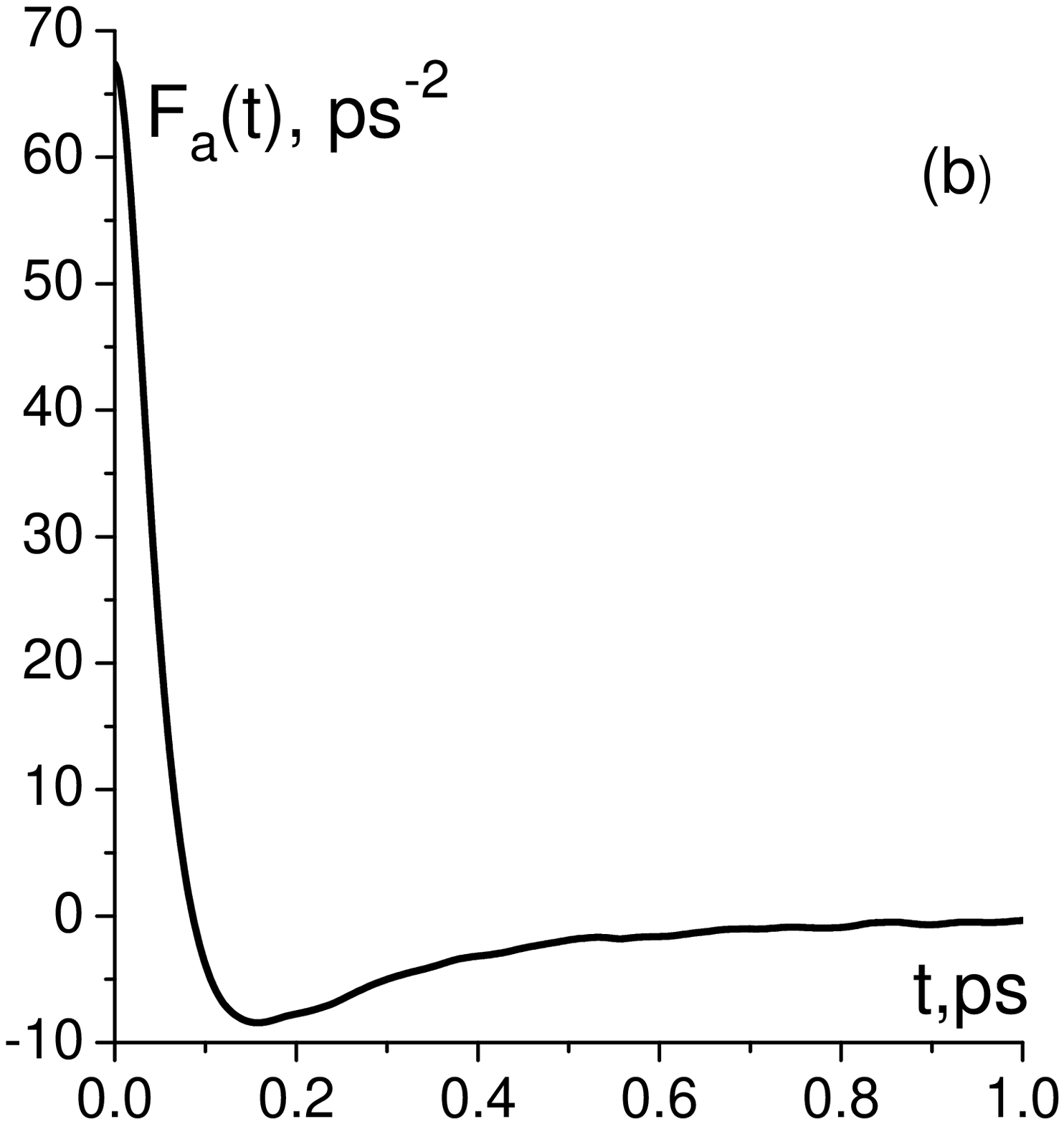}
  \includegraphics[height=0.179\textheight,angle=0]{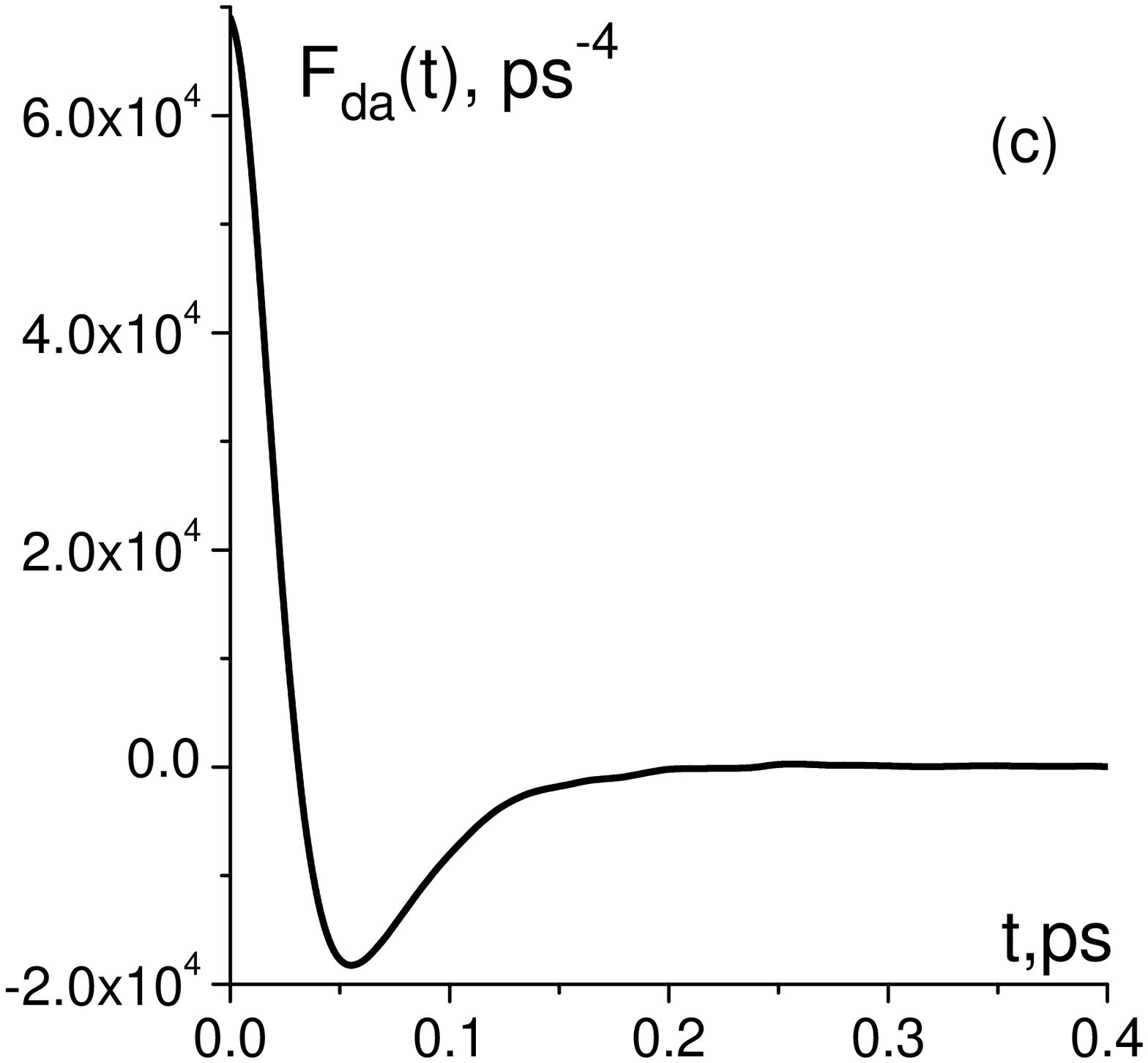}
   \includegraphics[height=0.179\textheight,angle=0]{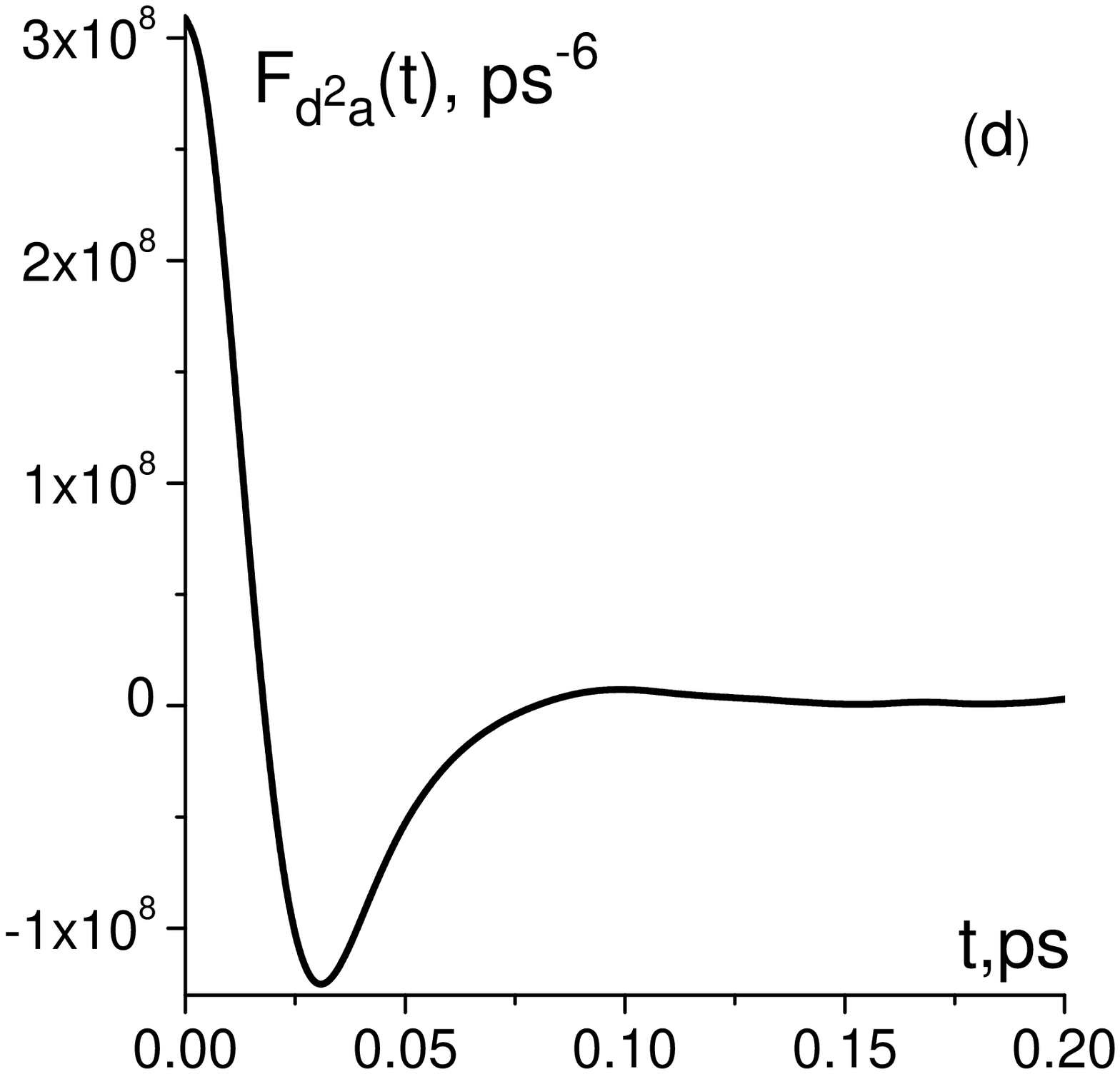}}
 \caption{Time correlation functions $F_v(t)=\langle \mathbf v_i \mathbf v_i(t)\rangle/\langle \mathbf v_i \mathbf v_i\rangle$ (a),
 $F_a(t)=\langle \mathbf a_i \mathbf a_i(t)\rangle/\langle \mathbf v_i \mathbf v_i\rangle$ (b),
 $F_{da}(t)=\langle \dot{\mathbf  a}_i \dot{\mathbf  a}_i(t)\rangle/\langle \mathbf v_i \mathbf v_i\rangle$ (c) and
$F_{d^2a}(t)=\langle \ddot{\mathbf  a}_i \ddot{\mathbf
a}_i(t)\rangle/\langle \mathbf v_i \mathbf v_i\rangle$  (d) of Ne
taken at the temperature $T=$295 K and densities $\rho=$2190
kg/m$^3$ (upper row), $\rho=$1400 kg/m$^3$ (middle row) and
$\rho=$691 kg/m$^3$ (lower row).}
\end{figure*}

Molecular dynamics simulations were performed on a system of 4000
particles interacting via pair Lennard-Jones potential with
parameters corresponding to Ne. The simulation setup and effective
potentials were essentially the same as in the previous study on
supercritical Ne \cite{Bry17}. We performed simulations for 12
densities in the range from 691 kg/m$^3$ to 2190 kg/m$^3$ of
supercritical Ne at temperature 295~K with the purpose to estimate
time evolution and calculate autocorrelations of particle
velocities, accelerations and their higher time derivatives. The
time step in simulations in the microcanonical ensemble was 0.5
fs, and the pressure calculated from MD simulations was in perfect
agreement with the available NIST data \cite{NIST} for densities
of supercritical Ne up to 1200 kg/m$^3$. The analytical
expressions for the acceleration and its derivatives,
\bea\label{microEqn}\nonumber && \mathbf a_i=\dot{\mathbf
v}_i=\sum\limits_{j\ne i=1}^N\mathbf a_{ij},\qquad \mathbf
a_{ij}=-\displaystyle\frac{1}{m}
\Phi'(r_{ij})\displaystyle\frac{\mathbf r_{ij}}{r_{ij}},\qquad
\Phi'(r)=\frac{d\Phi(r)}{dr},\\
&&\nonumber\dot{\mathbf
a}_i=\displaystyle\frac{1}{m}\sum\limits_{j\ne i=1}^N
\left\{-\Phi'(r_{ij})\frac{\mathbf
v_{ij}}{r_{ij}}-\left[r_{ij}\Phi''(r_{ij})-\Phi'(r_{ij})
 \right]\frac{\mathbf r_{ij}\left(\mathbf r_{ij}\cdot\mathbf v_{ij}
 \right)}{r_{ij}^3}
\right\},\\
&&\nonumber \ddot{\mathbf a}_i=-\frac{1}{m}\sum\limits_{j\ne
i=1}^N\left\{ \Phi'(r_{ij})\frac{\mathbf{a}_{ij}}{r_{ij}}+\left[
r_{ij}\Phi''(r_{ij})-\Phi'(r_{ij}) \right]\right.\\
&&\left.\times \left[\frac{2\mathbf v_{ij}(\mathbf
r_{ij}\cdot\mathbf v_{ij})+\mathbf r_{ij}\left(\mathbf
v_{ij}^2+(\mathbf r_{ij}\cdot\mathbf
a_{ij})\right)}{r_{ij}^3}-\frac{3\mathbf r_{ij}(\mathbf
r_{ij}\cdot\mathbf v_{ij})^2}{r_{ij}^5}
+\Phi'''(r_{ij})\frac{\mathbf r_{ij}(\mathbf
r_{ij}\cdot\mathbf v_{ij})^2}{r_{ij}^3} \right]\right\},\\
&&\nonumber\qquad\qquad\qquad\qquad\quad \mathbf r_{ij}=\mathbf
r_{i}-\mathbf r_{j},\quad \mathbf v_{ij}=\mathbf v_{i}-\mathbf
v_{j},\quad \mathbf a_{ij}=\mathbf a_{i}-\mathbf a_{j},\eea
involving the derivatives of Lennard-Jones potentials
$\Phi(r)\equiv\Phi_{LJ}(r)$, were used to estimate their time
evolution along particle trajectories in MD simulations.

To investigate the system dynamics in more detail, we do not limit
ourselves by the VAFs only and perform computer simulations for
the acceleration autocorrelation functions as well as for the
TCFs, defined on the higher derivatives of the velocity up to the
4-th order. Except for a purely theoretical interest, the obtained
results have an applied aspect too, providing us with the values
$\Gamma_j$ that are used in our method.

The results of MD simulations for some of the TCF of Ne are
presented in Fig.~1. It is seen that the VAF of the fluid at the
highest studied density $\rho$=2190 kg/m$^3$ has a pronounced
minimum at $t$=0.1 ps. This minimum is related to the
back-scattering of the fluid particle trapped in the cage, which
is formed by its neighborhood. At the intermediate times,
vibration of the particle causes a rearrangement of the solvation
shell, allowing the particle to travel away from its initial
position, until it is retrapped at $t$=0.25 ps by another cage
with much less local ordering of the particles. At larger times,
the particle escapes from all local traps, and the oscillations
become completely damped.

At the intermediate density $\rho$=1400 kg/m$^3$, the effects of
the local ordering of the molecules are almost negligible except
the time $t$=0.3 ps, where the corresponding VAF has an inflection
point. At the density $\rho$=691 kg/m$^3$, the molecules packing
is too small to produce any sign of the local ordering, which
would be able to trap the particle, and the VAF smoothly decays on
the whole time domain.

The results of particle dynamics in terms of the VAFs are
interesting to compare with those of acceleration autocorrelation
functions $F_a(t)$, depicted in the column b) of Fig.~1. At the
density $\rho$=2190 kg/m$^3$, this function changes its sign in
the time interval 0.04$\div 0.14$~ps. It means that the force,
acting on the Ne particle, reverses its direction due to the
particle scattering on surrounding molecules. Thus, the interval
about 0.05~ps can be associated with the effective time between
collisions.

On the other hand, the depth of the well in the acceleration
autocorrelation function strongly decreases with the density
reduction. This phenomenon can be explained in the following way:
at low densities the particle changes its direction of motion due
to the multiple low-angle scattering whereas at high density the
``head-on'' collisions dominate, yielding the profound minimum in
$F_a(t)$.

The last two columns in Fig.~1 show that autocorrelation
functions, associated with higher derivatives of force, decay much
faster than $F_a(t)$, while locations of the minima cease to
depend on the system density. It is quite expected, since the rate
of the force change is defined by a continuous interaction of the
tagged particle with its surrounding rather than by an effective
(density dependent) collision mechanism between the molecules. The
same is also true for the auto\-correlation functions $F_{d^2
a}(t)$, while their relaxation times are shorter as compared with
those of $F_a(t)$, since the former TCFs are constructed on much
faster dynamical variables. The auto\-correlation functions, dealt
with the third derivative of the force, bring no essential physics
to understanding of the fluid dynamics, and, consequently, are not
presented in Fig.~1.

The evaluated autocorrelation functions allow us to obtain the
values of $\Gamma_s$, which are the input parameters in our
theory. Indeed, zero time values of the autocorrelation functions
$F_a(t)$, $F_{da}(t)$, $F_{d^2a}(t)$ define, correspondingly, the
2-nd, the 4-th and the 6-th frequency moments of the VAF,
\bea\label{W24} \langle\omega^2\rangle=\frac{\langle\mathbf a_i
\mathbf a_i\rangle}{\langle\mathbf v_i\mathbf v_i\rangle},\qquad
\langle\omega^4\rangle=\displaystyle\frac{\langle\dot{\mathbf
a}_i\dot{\mathbf a}_i\rangle}{\langle\mathbf v_i\mathbf
v_i\rangle},\qquad
\langle\omega^6\rangle=\displaystyle\frac{\langle\ddot{\mathbf
a}_i\ddot{\mathbf a}_i\rangle}{\langle\mathbf v_i\mathbf
v_i\rangle},
 \eea
which can be calculated directly via the positions $\mathbf r_i$
and velocities $\mathbf v_i$ of all the particles in computer
simulations using the microscopical equations of motion
(\ref{microEqn}). The expressions for higher order frequency
moments can be obtained straightforwardly. In our study, we
calculate frequency moments up to the 8-th order directly by MD,
and only the 10-th order frequency moments are evaluated
approximately by taking the second order derivative of the
autocorrelation functions $\langle \dddot{\mathbf{a}}_i
\dddot{\mathbf{a}}_i(t)\rangle/\langle \mathbf{v}_i
\mathbf{v}_i\rangle$ at $t=0$.

The above mentioned frequency moments can be connected by the
relations
\bea\label{WvsGamma}\nonumber
\Gamma_1=\langle\omega^2\rangle,\qquad\Gamma_2=\displaystyle\frac{\langle\omega^4\rangle}{\langle\omega^2\rangle}-\langle\omega^2\rangle,\qquad
\Gamma_3=\displaystyle\frac{\langle\omega^6\rangle\langle\omega^2\rangle-\langle\omega^4\rangle^2}{\langle\omega^4\rangle\langle\omega^2\rangle-
\langle\omega^2\rangle^3},
\\
\Gamma_4=\frac{1}{\Gamma_1\Gamma_2\Gamma_3}\left\{\langle\omega^8\rangle-\Gamma_1\left[\left(\Gamma_1+\Gamma_2
\right)^3+2\Gamma_2\Gamma_3\left(
\Gamma_1+\Gamma_2\right)+\Gamma_2\Gamma_3^2 \right] \right\}
 \eea
 with the SCFs $\Gamma_j$,
constructed on the orthogonal set of the dynamic variables. The
expression for the SCFs $\Gamma_j$ with higher indexes $j$ can be
evaluated straightforwardly using certain recurrent relations
(see, for instance, Eq.~(4) of Ref.~\cite{Bafile1}).

\section{Time behavior of the VAFs: theory\label{secV}}

Expressions like (\ref{spectr2})-(\ref{spectrM2}) and
(\ref{spectr3})-(\ref{spectrM3}) allow us to study the frequency
dependence of SFs in the broad domain of parameters $\Gamma_j$ as
it has been discussed in Sec.~III. It has been shown in
Ref.~\cite{CMPour} that the VAFs, obtained within our approach,
have the same frequency moments up to the order $(2s-2)$ as their
counterparts, obtained within the MA \cite{GCM3,GCM4}.
What is the most essential, our approach reproduces
also the exact result for the next frequency moment, $\langle
\omega^{2s}\rangle=\frac{1}{2\pi}\int_{-\omega_c}^{\omega_c}\omega^{2s}\tilde
F_s(\omega)d\omega$, that can be easily verified, for instance, by
a direct integration of the corresponding SFs using
Eqs.~(\ref{spectr2})-(\ref{spectrM2}) and
(\ref{spectr3})-(\ref{spectrM3}). In other words, we have an extra
(correct) term $\sim t^{2s}$ in the Taylor expansion series of the
VAFs as compared with the MA case at the same number of the input
parameters $\Gamma_j$. This allows us to consider the method based
on the effective summation of the infinite continued fractions as
more efficient at the description of the VAFs behavior not only at
the small but also at the intermediate times. Besides, the
square-root dependence (\ref{sqrtEps}) of the SFs causes the
long-time tails appearance, though in a different manner as it is
usually predicted in the MCT \cite{MCT1,MCT2,MCT3}.

To start with an investigation of the VAFs dynamics, we present in
Table~I the values of the lowest order SCFs $\Gamma_1$ and ratios
$\Gamma_{j+1}/\Gamma_j$ at various densities. It is evident from
the Table~I that the above mentioned ratio tends noticeably to the
unity with the increase of the hierarchy level $j$. Hence, our
basic assumption about the fast convergence of the SCFs $\Gamma_j$
looks well-grounded.

\begin{table}[htb]
\caption{Values of the SCFs $\Gamma_1$ and ratios
$\Gamma_{j+1}/\Gamma_j$ at the highest, intermediate and lowest
densities studied.}
\begin{center}
\begin{ruledtabular}\begin{tabular}{cccccc}
 \,\,\,&$\Gamma_1$, ps$^{-2}$ & $\Gamma_2/\Gamma_1$ & $\Gamma_3/\Gamma_2$ &
  $\Gamma_4/\Gamma_3$ &  $\Gamma_5/\Gamma_4$\\
     \hline
$\rho=2190$ kg/m$^3$ &747.9  &1.774&3.606&2.406&1.45\\
$\rho=1400$ kg/m$^3$ &233.611 &4.531&3.779&2.473&1.9\\
$\rho=691$ kg/m$^3$  &67.385 &14.377&3.843&2.521&1.665\\
 \end{tabular}\end{ruledtabular}
     \label{ratioGamma}
     \end{center}
     \vspace{-2mm}
     \end{table}

A similar convergence tendency can be traced for the lowest order
relaxation time $\tau_1=1/\tilde{\phi}_1(0)$, evaluated at various
indexes $s=1\div 5$ of the hierarchy level as it is clearly
visible from Fig.~2.
\begin{figure}[htb]
\centerline{\includegraphics[height=0.25\textheight,angle=0]{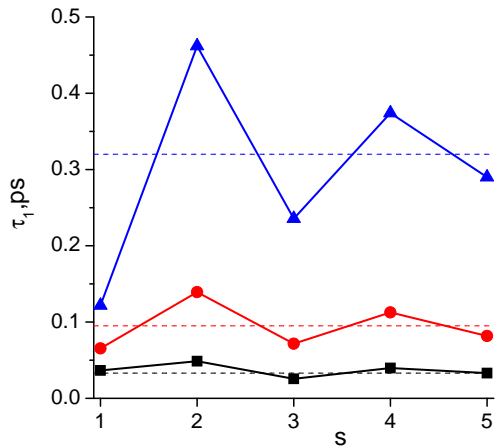}}
 \caption{Dependence of the lowest order relaxation time $\tau_1$ on the hierarchy level index $s$ at the fluid densities $\rho$=691 kg/m$^3$
 (triangles), 1400 kg/m$^3$ (circles), and 2190 kg/m$^3$ (squares).
The dashed lines correspond to the values taken from the computer
simulations. The straight lines are a guide for eye.}
\end{figure}
\noindent It is seen that $\tau_1$ oscillate around
their values attainable at some large $s\gg 1$, approaching them
from above (even $s$) or below (odd $s$). A convergence grows
rapidly with the density increase. Besides, the values of $\tau_1$
decrease with density, since the relaxation times of the memory
kernels are lower in the dense fluids.

\begin{figure*}[htb]
\centerline{\includegraphics[height=0.22\textheight,angle=0]{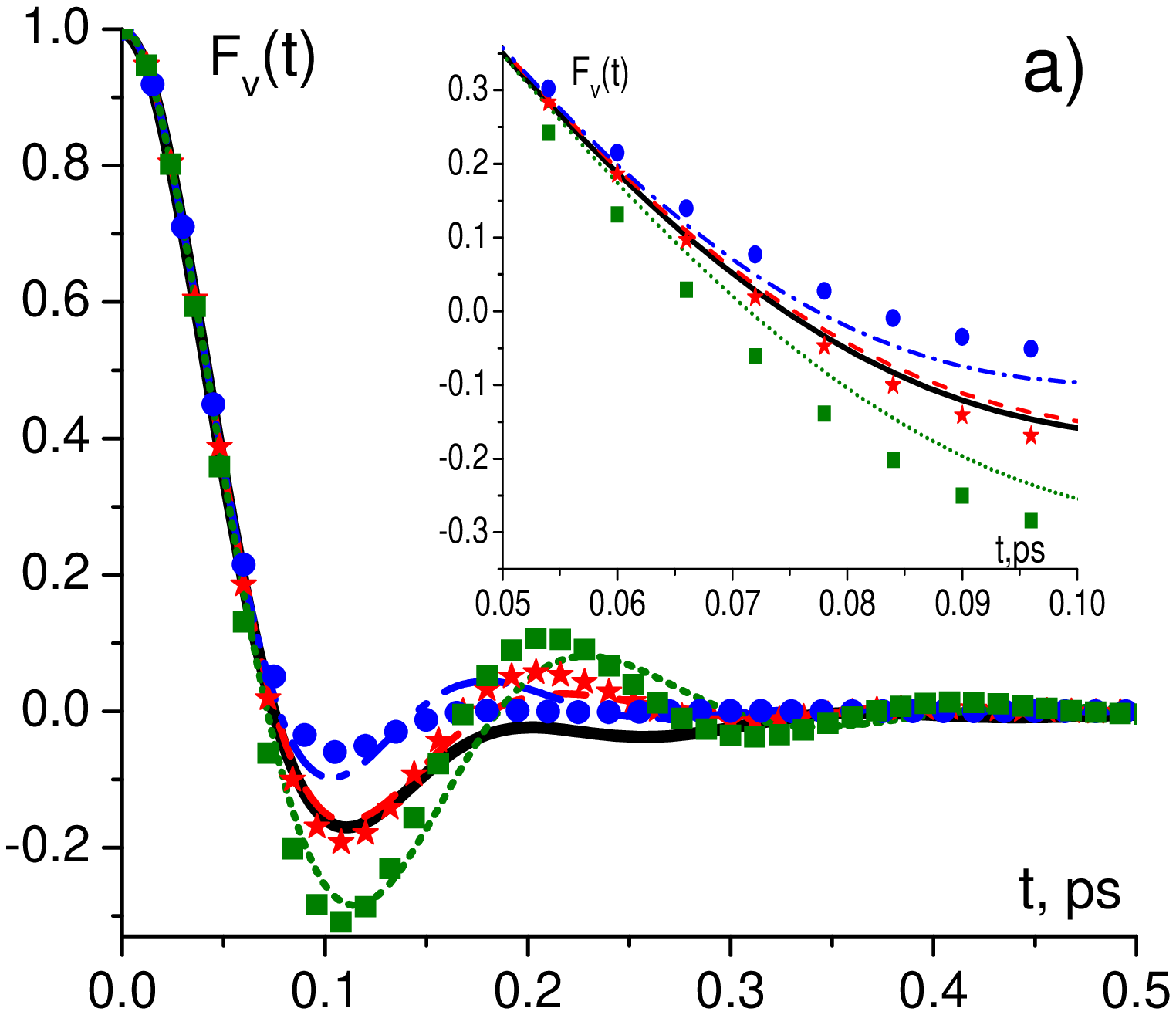}
 \includegraphics[height=0.22\textheight,angle=0]{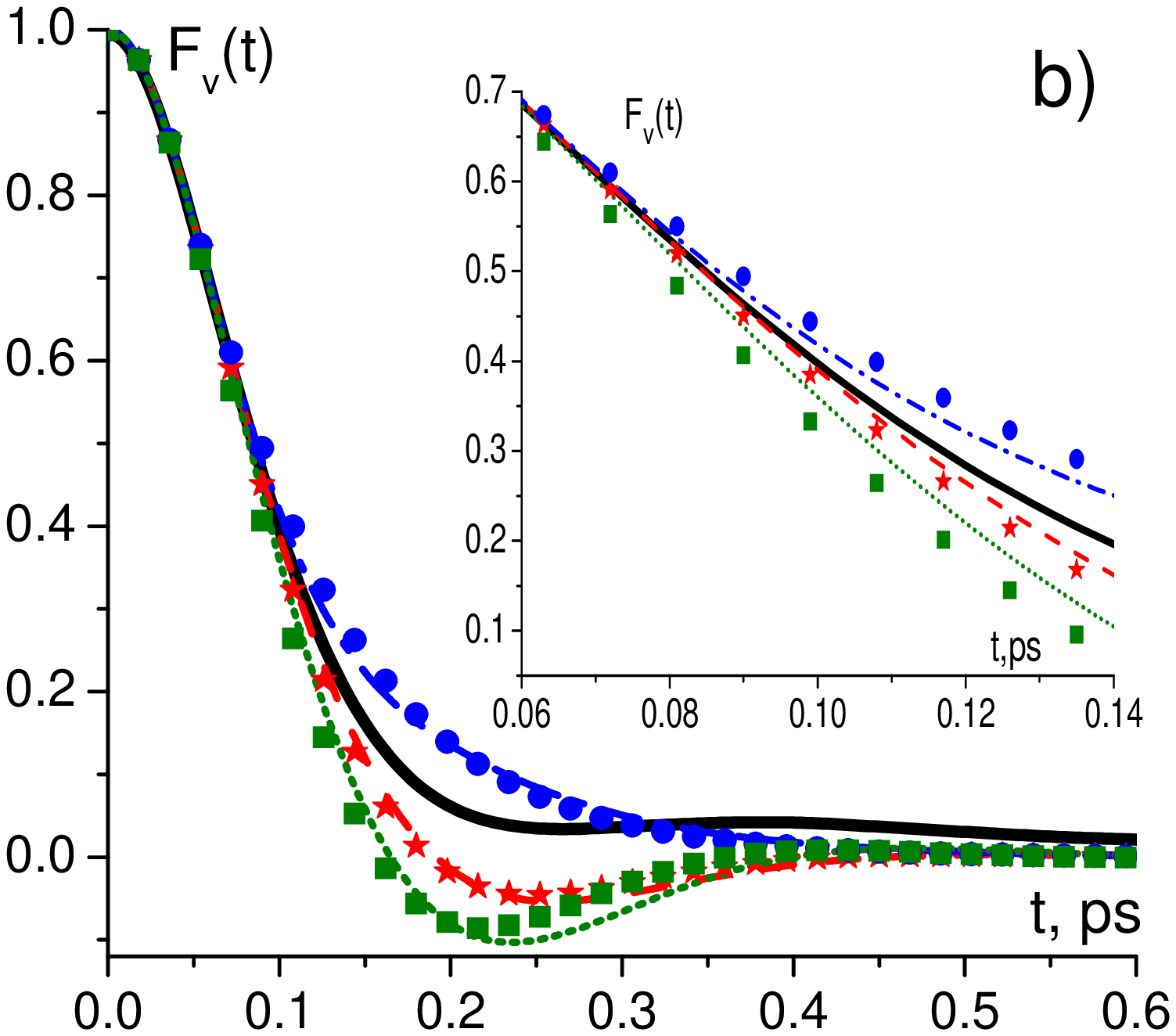}
  \includegraphics[height=0.22\textheight,angle=0]{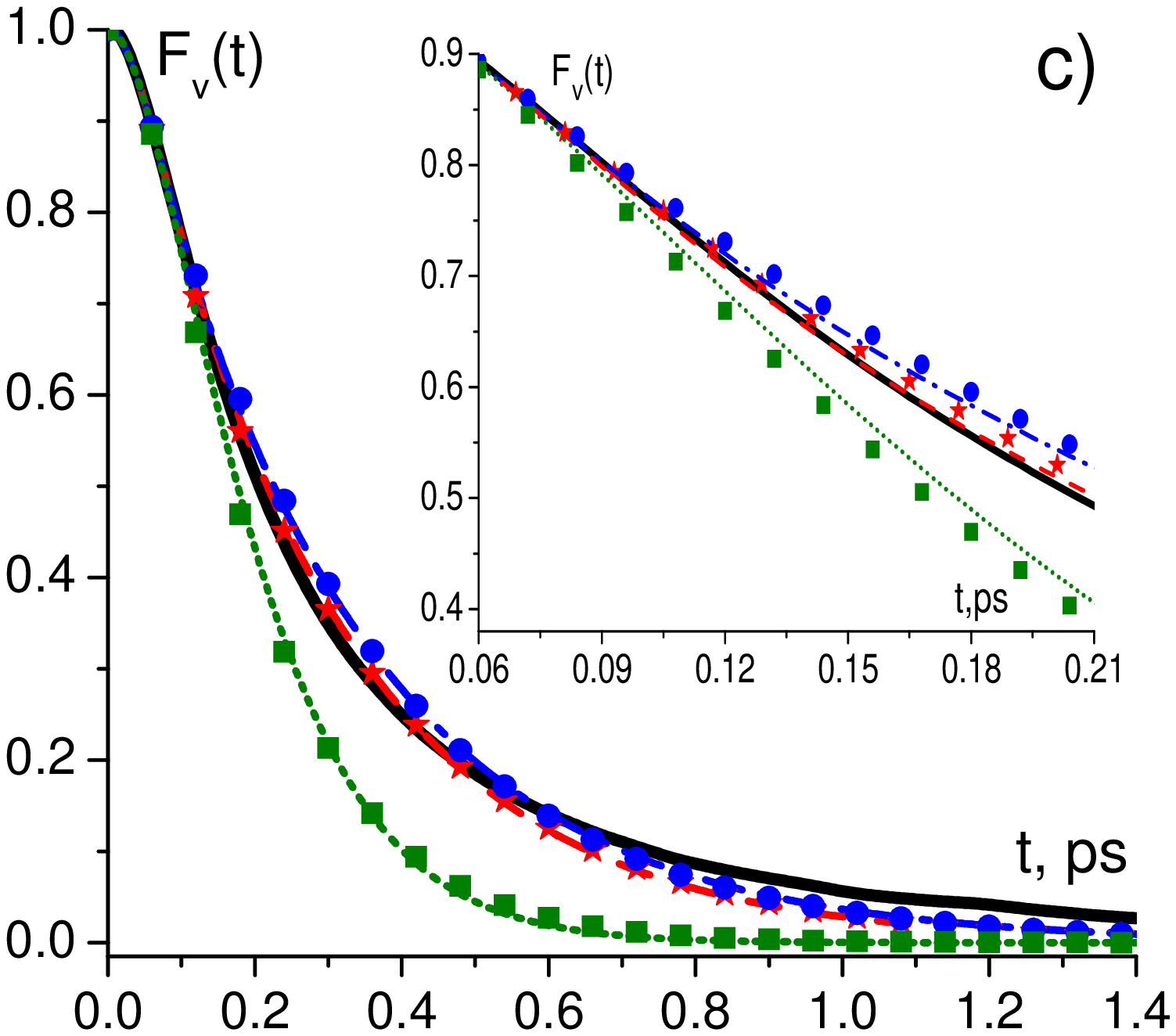}}
 \caption{VAFs of the liquid Ne at the densities $\rho=$2190 kg/m$^3$ (a), 1400 kg/m$^3$ (b)
 and 691 kg/m$^3$ (c). The solid lines correspond to the MD
 data. Other lines correspond to the VAFs, obtained by an
 effective summation of the infinite continued fractions
up to the 3-rd (dotted
 line), 4-th (dashed-dotted line) and 5-th (dashed
 line) hierarchy levels. The symbols correspond to the VAFs, obtained
 within MA3 (squares), MA4 (bullets) and MA5 (stars) approximations. In
the insets: time behavior of the corresponding VAFs at small $t$.}
\end{figure*}

Thus, having the values of all the input parameters $\Gamma_j$,
$j=\{1,\ldots,5\}$, and performing the inverse Fourier
transformation of the corresponding SFs, which can be obtained in
a similar way as those in Sec.~III, we are able to calculate the
velocity autocorrelation functions at different levels of
approximation. The results for the VAFs of liquid Ne at various
densities, calculated directly by MD simulations, as well as those
obtained analytically at different approaches are presented in
Fig.~3.

One can draw several conclusions when inspecting Fig.~3. First of
all, a convergence of all results to the MD data (denoted by the
solid master curve) with increasing of the approximation order is
clearly noticeable. Both lines (corresponding to our approach) and
symbols (related to the modes approximation) follow the solid
black line the better, the higher $\Gamma_s$ are taken into
account. It is quite expected, since by increasing the hierarchy
level $s$ we ensure more frequency moments of the VAFs to be
satisfied.

Secondly, as we have mentioned at the beginning of this Section,
the effective summation of the infinite continued fractions
ensures more terms in time series of the VAFs that reproduce the
exact fluid dynamics as compared with the fractions truncation
(which has been shown to be equivalent to the mode approximation).
Indeed, it is seen in the insets of Fig.~3 that for all the
densities the curves, correspond\-ing to the fixed level of
approximation $s$, leave the so\-lid master curve later than their
symbolic counterparts.

Thirdly, and what is the most surprising, the dashed curve at the
highest reported density coincides with the solid one in a rather
broad time interval, including that of the minimum of VAF, related
to the cage effect. Moreover, the difference between the results,
obtained in our approach (lines), and those, obtained by a
truncation of the continued fractions (symbols), is quite
noticeable. At the lowest density, this difference almost
vanishes, and a question, which method should be chosen to
describe the fluid dynamics, becomes less relevant.

Now one remark is to the point. It is possible to calculate the
generalized memory kernels of different orders using the MD data
and compare them with those, evaluated by an effective summation
of the continued fractions or by the MA. The point is that the
autocorrelation functions of different orders obey the recurrence
relations \cite{Bafile1}, allowing one to express the generalized
kinetic kernels $\tilde{\phi}_j(z)$ via SCFs and Laplace
transforms of the TCFs (see, for instance, Eqs.~(6)-(9b) of the
cited paper). However, to evaluate the dynamics of
kernels at the same accuracy level as it has been done in the case
of VAFs (Fig.3), one has to possess (at least) the same number of
SCFs $\Gamma_j$, which should be of higher orders as compared with
those used in our study. In particular, to calculate reliably the
1-st order kernel we need the 6-th order SCF and so on. Since, in
any case, we have a limited number of parameters, obtained from
the MD calculations, a dynamics of the memory functions
$\tilde{\phi}_j(z)$ (even started from $j=1$) would be obtained
with worse accuracy than that of VAFs.

As we have already mentioned, the long time behavior of the TCFs
remains a topical problem in the physics of fluids
\cite{MCT1,MCT2,MCT3,long_tail1,Ald70}. In our recent paper
\cite{CMPour}, we have paid much attention to the studies of the
long-time tails formation in fluids. In particular, we obtained
quite simple expressions for the transition time $\tau_H$ to the
hydrodynamic regime at the assumption that all the $\Gamma$
converge to their asymptotic value starting from $s=2$. We have
shown that there could be not only the simple poles of the
corresponding SFs, but also another kind of discontinuity, like
the essential singularity points. Since the kinds of the
singularities and the localization of the discontinuities on the
complex frequency plane are known to define
\cite{russians1,russians2} the values of $\tau_H$, this fact
brings new features into the long time dynamics of the VAFs,
renormalizing the corresponding transition times.

A similar attempt to express the transition times analytically can
be performed at the approximation of the $s$-th hierarchy level.
However, it is much more instructive to estimate $\tau_H$
numerically at several thermodynamic points, taking the values of
$\Gamma_j$, obtained by computer simulations, as the input
parameters.

\begin{figure}[htb]
\centerline{\includegraphics[height=0.25\textheight,angle=0]{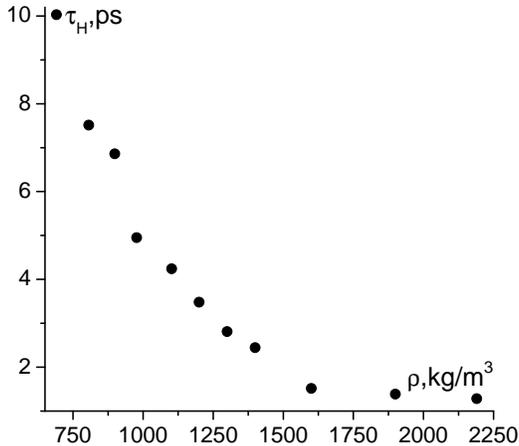}}
 \caption{Dependence of the transition time $\tau_H$ to the hydrodynamic regime on the fluid density.}
\end{figure}

In Fig.~4 we present the density dependence of the time $\tau_H$,
needed for the fluid to reach the hydrodynamic stage of evolution.
In our calculation, we have performed an effective summation of
the infinite continued fraction within the approach, described in
Sec.~II, and used the corresponding SCFs up to the order $s=4$,
for which the direct MD data are available. The transition time
was estimated by the contact point of the envelope curve $a
t^{-3/2}$ and the corresponding VAF.

It is seen from Fig.~4 that the transition time weakly depends on
$\rho$ at the superdense states of the fluid. Then, starting from
$\rho<1500$ kg/m$^3$, it grows by almost an order of magnitude at
the smallest density studied. If we recall a concept of the
effective time between particle collisions, which can be related
to the instant, when the ``force--force'' autocorrelation function
changes its sign (see the column (b) of Fig.~1), it becomes
evident, that from 30 (at the highest density) to about 100 (at
the lowest density) particle collisions are required to form the
long-time tails in the VAFs. In this context, our results agree
with those, obtained in Ref.~\cite{Bafile2}, where the number of
particle collisions prior to the long-time tails appearance was
estimated by the ratio of $\tau_H$ and the Enskog mean time
$\tau_E$ between collisions for the equivalent hard spheres fluid.
On the other hand, the transition time at $\rho\sim 1500$ kg/m$^3$
and $T=295$ K (that correspond to the dimensionless density
$\rho\sigma_{LJ}^3=1$ and temperature $k_B T/\varepsilon_{LJ}=8$),
being expressed in the dimensionless time units
$\tau_H/\tau_{LJ}$,
$\tau_{LJ}=\sigma_{LJ}\sqrt{m/\varepsilon_{LJ}}$, is close to the
MD data of Ref.~\cite{russiansJCP} (look at the vicinity of the
first bullet in the inset in Fig.~3(c) of the cited paper).

As we have already mentioned, our approach can be quite promising
at the study of the crossover to the Navier-Stokes hydrodynamics,
since it provides a description of the SFs topology in the complex
frequency plane as well as a systematization of the singularities
types \cite{russians1,russians2}, whereas the MCT does not give a
direct recipe for estimation of this value. However, in contrast
to the MCT, the above presented scheme does not allow to calculate
reliably, for instance, the long tail amplitude $a$. Anyway, we do
not expect our results to agree quite satisfactory with those
predicted by the MCT, since these two methods rest upon different
basic points.

In our theory, we do not operate explicitly by any dissipative
characteristics of the system, originating from the
hydrodynamically based MCT approach, like the viscosity  or
self-diffusion coefficient (though the last one can be expressed
by the corresponding spectral functions taken at zero frequency,
see Sec.~III). The fluid dynamics is completely defined by the
static characteristics of the system given by the parameters
$\Gamma_j$, and the obtained results well agree with the computer
simulations data in the quite broad time domain of the system
evolution.

In the framework of MCT, a basic set of the dynamic variables
consists of higher products \cite{MCT3} rather than higher
derivatives of the densities of conserved quantities.
In the MCT approach \cite{MCT1,MCT3}, the system
dynamics is well described at the long times, but this method does
not allow to obtain the reliable results at other timescales (in
particular, the results obtained within MCT based methods do not
obey the sum rules).

To conclude the discussion about an interrelation of our method
and the MCT, we would like to mention some important points. To a
certain extent, our theoretical scheme can formalize the idea of a
self-consistent description of the density fluctuations in a
similar manner as this is done by the MCT, where the description
of the system dynamics is made by the closed integro-differential
equations \cite{Reichman2}.  Indeed, Eqs. (\ref{Eqns-time}) are of
the similar structure as those considered in the framework of MCT:
we can treat the relation (\ref{phiS-1}) as an ansatz which allows
to express the highest order memory function via the $(s-1)$-th
order kernel to close the chain of equations.

What is more interesting, the above mentioned formal similarity
between our method and the MCT does not end at this point. There
is one more common feature of both approaches: a non-analytic
dependence of the SFs on frequency. However, in the MCT framework
it occurs in the zero frequency domain \cite{long_tail1}, while in
our approach the corresponding SFs depend non-analytically on
$\omega$ in the vicinity of the cut-off frequency $\omega_c$, see
Eq.~(\ref{sqrtEps}). The coefficient at the square root in
(\ref{sqrtEps}) defines the value of the long tail amplitude at
the accepted condition of the 2-nd order hierarchy.

However, as we have already mentioned, this value would differ
from the prediction of the MCT. Moreover, one should speak about a
certain envelope curve rather than the long-time tail $at^{-3/2}$.
The point is that our assumption about very rapid (basically,
instantaneous) convergence of the SCFs $\Gamma_j$ to their
asymptotic limit is rather oversimplified. Actually, one deals
with a much slower convergence of the higher order relaxation
times with $j$. The above mentioned
oversimplification leads to a pronounced modulation of the
obtained long-time tails, as discussed in detail in our recent
paper \cite{CMPour}.

One of the possible ways to diminish these oscillations consists
in a construction of some kinds of the periodic continued
fractions according to the following ansatz: \bea\label{approxSn}
\tilde{\phi}_s(z)\approx\tilde{\phi}_{s+n}(z),\qquad n\ge 2. \eea
From the physical point of view, it corresponds to a more
realistic approximation for the VAFs, when the high order
relaxation times approach their asymptotic values, alternately,
from below or above \cite{our_assump2} rather than instantly, as
it happens within our scheme.

From the mathematical viewpoint, we obtain a bilinear form with
respect to the highest order memory kernel $\tilde{\phi}_s(z)$,
like it was in the case of our basic approximation
(\ref{approxS}). However, a discriminant of the quadratic equation
would have a much more complicated structure, being the $2n$-th
order polynomial in $\omega$. Zeros of the above mentioned
polynomial define the cut-off frequencies $\omega_{c i}$,
$i=\{1,\ldots,n\}$ and, consequently, the gaps in the SFs. With
increasing of $n$, the interval of the frequencies $[\omega_{c 1},
\omega_{c n}]$ broadens; at the same time, the lowest cut-off
frequency $\omega_{c 1}$ moves to the left, and the highest
cut-off frequency $\omega_{c n}$ shifts to the right. The
frequency behavior of the SFs at the gaps' edges is very similar
to (\ref{sqrtEps}), being proportional to
$\sqrt{|\omega-\omega_{ci}|}$. The coefficients of the
proportionality (long tails amplitudes $a_i$) can be ordered by
their magnitudes, $|a_1|>|a_2|>\ldots|a_n|$. Due to the low
frequency shift of the $\omega_{c i}$ at small $i$ and to a
negligible contribution of the terms with large $i$, the net
effect would be a smoothing of the oscillations in the long-time
tails of the VAFs. It is clear that a construction of such
periodic continued fractions has to be well grounded from the
physical viewpoint (in the first turn, due to requirement of the
sum rules \cite{CMPour}), and could be a subject of the separate
studies.

\section{Conclusions\label{secVI}}

In this paper, we propose a simple ansatz for the high order
kinetic kernels of the fluid that allows us to model the system
dynamics on a quite broad time interval. Our approach is based on
the physically grounded assumption about convergence of the
relaxation times of the high order memory functions
\cite{our_assump2,CMPour}, which have a purely kinetic origin. As
a result, the infinite continued fractions for the Laplace
transforms of the VAFs can be effectively summed up, yielding the
closed form that depend only on limited number of the SCFs, which
can be taken from computer simulations.

To a certain extent, such an approach resembles the GCM formalism
\cite{GCM1,GCM2,GCM3,GCM4}, which also assumes a description of
the system by a finite set of the parameters (the static
correlation functions and corresponding relaxation times).
However, the distinctive feature of our approach is that, in
contrast to the GCM theory (or interrelated with it the MA
formalism), there is an intrinsic frequency dispersion in the
memory kernels of all orders.

The above mentioned dispersion turns out to have a double impact
on the fluid dynamics. First of all, it provides a larger number
of terms  in the time series of the VAFs that agree with the MD
simulations data, allowing one to describe the system dynamics at
short and intermediate times more precisely as compared with the
GCM (MA) at the same level $s$ of the hierarchy. Quite
unexpectedly, our approach was found to be especially promising at
the description of the super dense fluids, while at the lower
densities the difference from the results, obtained within MA
framework, becomes less evident, and the issue which method is
more preferable becomes less relevant.

Secondly, unlike the GCM, an application of our method reaches far
beyond this time domain, making it possible to estimate the
transition times from the kinetic to the hydrodynamic stages of
evolution, and even to study the processes of long-time tails
formation (though, at least in its present form, our approach
cannot compete with the MCT in the description of late stages of
the system dynamics). In this context, our method has some common
features with the so-called ``combined'' ones
\cite{russians1,russians2}, based on investigation of the
discontinuities of some (properly constructed) SFs on the complex
frequency plane. It is shown that the transition time increases
considerably with the density decrease. It requires some tens of
the (effective) particles collisions for the long-time tails
formation that agrees completely with results of
Refs.\cite{MHforMemKer,Bafile2,russiansJCP}. At this issue, our
result is also consistent with the MCT viewpoint
\cite{Ald70,Reichman2}, which considers the sequence of correlated
collisions as a microscopic precursor of the long-time tails
appearance.

An occurrence of the cut-off frequency $\omega_c$, which modulates
the power law dynamics of the fluid at long times, can be
explained in terms of the generalized Green-Kubo relations
\cite{MorozovBook,Boo}. There is a narrowing of the dissipation
channel in the high frequency domain \cite{CMPour}, which cannot
compensate a persistent renormalization of the vibrational modes
by a non-vanishing imaginary part of the generalized
self-diffusion coefficient.

It should be emphasized that a more accurate calculation of the
high order memory kernels, presumably based on the infinite
periodic continued fractions, is strongly desirable, because they
are expected to ensure a smoother relaxation of the corresponding
VAFs at long times, and to bring the long-time characteristics of
the fluid much closer to those of the MCT prediction. Such a
method requires not only a greater number of the relevant SCFs,
which have to be taken from computer simulations, but also implies
a very precise design of the higher order kinetic kernels for all
the necessary sum rules to be satisfied. We believe that this task
is quite promising and worthy the efforts to be spent to solve the
problem how to describe the many-particle system dynamics on the
whole time domain in the framework of a unified self-consistent
approach.

\section*{Acknowledgement}
This study was partially supported within the project of the
European Unions Horizon 2020 research and innovation programme
under the Marie Sk\l\,\!odowska-Curie grant agreement No 734276.

\end{document}